\documentclass[12pt,a4paper]{article}
\pdfoutput=1
\usepackage{jheppub}
\usepackage{setspace,verbatim} 
\usepackage{amsmath,amsfonts,amssymb,amsthm,graphicx,color}
\usepackage{verbatim}
\usepackage{enumerate}
\numberwithin{equation}{section}
\usepackage{bbm,bm}
\usepackage{braket,mathrsfs}
\usepackage{xcolor}
\usepackage{framed}
\usepackage{tikz}
\usetikzlibrary{decorations.pathmorphing,arrows.meta,bending,snakes}
\usepackage{caption}
\usepackage{subcaption}
\usepackage{dsfont}

\usepackage[normalem]{ulem}
\newcommand{\stkout}[1]{\ifmmode\text{\sout{\ensuremath{#1}}}\else\sout{#1}\fi}
\definecolor{darkgreen}{rgb}{0,0.4,0}
\definecolor{darkred}{rgb}{0.4,0,0}
\definecolor{darkblue}{rgb}{0,0,0.4}

\def\undertilde#1{\mathord{\vtop{\ialign{##\crcr
$\hfil\displaystyle{#1}\hfil$\crcr\noalign{\kern1.5pt\nointerlineskip}
$\hfil\tilde{}\hfil$\crcr\noalign{\kern1.5pt}}}}}

\usepackage[framemethod=TikZ]{mdframed}
\definecolor{darkcerulean}{rgb}{0.03, 0.27, 0.49}
	
\newcounter{ex}[section]
\newenvironment{ex}[3][]{%
\refstepcounter{ex}%
\ifstrempty{#2}%
{\mdfsetup{%
frametitle={%
\tikz[baseline=(current bounding box.east),outer sep=0pt]
\node[anchor=east,rectangle,fill=white]
{\strut };}}
}%
{\mdfsetup{%
frametitle={%
\tikz[baseline=(current bounding box.east),outer sep=0pt]
\node[anchor=east,rectangle,fill=darkcerulean!12]
{ #1:{\normalfont ~#2} };}}%
}%
\mdfsetup{innertopmargin=9pt,linecolor=darkcerulean,%
linewidth=1.2pt,topline=true,%
frametitleaboveskip=\dimexpr-\ht\strutbox\relax
}
\begin{mdframed}[]\relax%
\label{#3}}{\end{mdframed}}

\makeatletter
\def\@xfootnote[#1]{%
  \protected@xdef\@thefnmark{#1}%
  \@footnotemark\@footnotetext}
\makeatother

\begin{document}
\title{Bound on asymptotics of magnitude of three point coefficients in 2D CFT}
\author[1,2]{Sridip Pal}

\affiliation[1]{Department of Physics, 
University of California, San Diego\\
La Jolla, CA 92093, USA}

\affiliation[2]{School of Natural Sciences, 
Institute for Advanced Study, Einstein Drive\\
Princeton, NJ 08540, USA}

\emailAdd{sridip@ias.edu}

\abstract
{We use methods inspired from complex Tauberian theorems to  make progress in understanding the asymptotic behavior of the magnitude of heavy-light-heavy three point coefficients rigorously. The conditions and the precise sense of averaging, which can lead to exponential suppression of such coefficients are investigated. We derive various bounds for the typical average value of the magnitude of heavy-light-heavy three point coefficients and verify them numerically.}

\maketitle
%



\section{The Premise and the Result}
The modular invariance plays a pivotal role in constraining the data of two dimensional conformal field theory. In two dimensions, a conformal field theory can be consistently defined on any Riemann surface, in particular, on a torus. The shape of the torus is determined by modular parameter $\tau \in \mathbb{H}$, where $\mathbb{H}$ is the upper half plane. The modular transformation acts on $\tau$ and maps it to another point in $\mathbb{H}$, nonetheless the partition function of the conformal field theory defined on the torus remains invariant under such transformation. Physically, one cycle of the torus is interpreted as the spatial cycle while the other one is the thermal cycle. Modular transformation, for example, exchanges these cycles and thus can relate the low temperature behavior of a CFT with its high temperature behavior. This is how the universality in the low temperature behavior translates into a universal high temperature behavior, which is controlled by the asymptotic data of the CFT. Hence, one can get insight to the asymptotic behavior of the CFT data utilizing the modular invariance\cite{cardy1986operator,Hellerman:2009bu,KM,dattadaspal,HKS,dyer,Das:2017cnv,Collier:2016cls,Collier:2018exn,Cho:2017fzo,Bae:2017kcl,Kim:2019yrz,Kusuki:2018wpa,Kusuki:2019gjs,Benjamin:2019stq,Maxfield:2019hdt}. Using AdS-CFT correspondence, the asymptotic data translates to a statement about gravity in AdS$_3$, in particular black holes \cite{KM,dattadaspal,HKS,Das:2017cnv}. Having a precise and rigorous statement about asymptotics is of paramount importance in this regard. For example, one can see in \cite{Baur, HKS}, the sparseness criterion came out as a natural consequence of the CFT being holographic and having a Cardy like regime for density of states. This has only been achieved by being very precise about the distinction and similarity between $\Delta\simeq c\to\infty$ limit and $\Delta\to\infty$ with $c$ being finite limit.\\

\begin{figure}[!ht]
\centering
\includegraphics[scale=0.7]{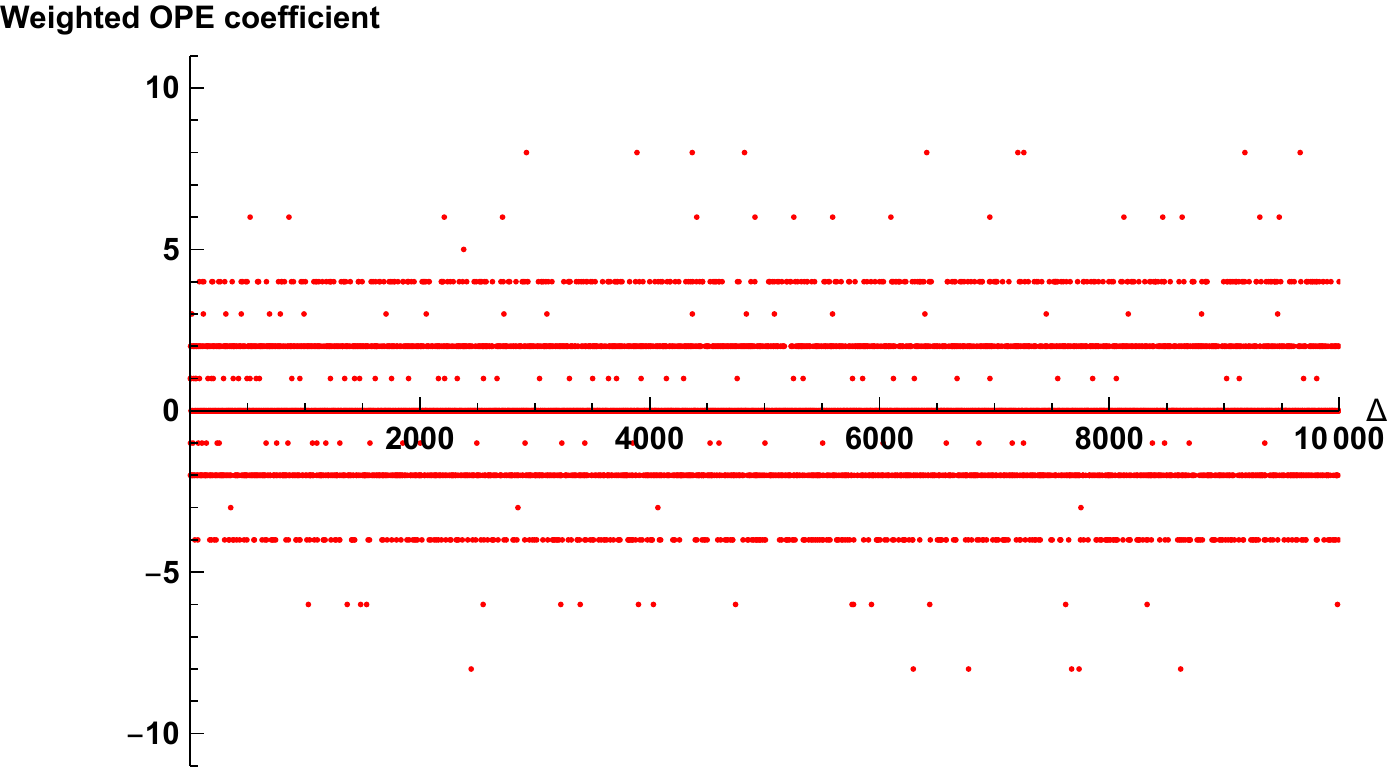}
\caption{$a_\Delta$, the $q$ expansion coefficient of the torus one point function of $\epsilon$ as a function of $\Delta$.}
\label{fig:eta}
\end{figure}

\begin{figure}[!ht]
\centering
\includegraphics[scale=0.5]{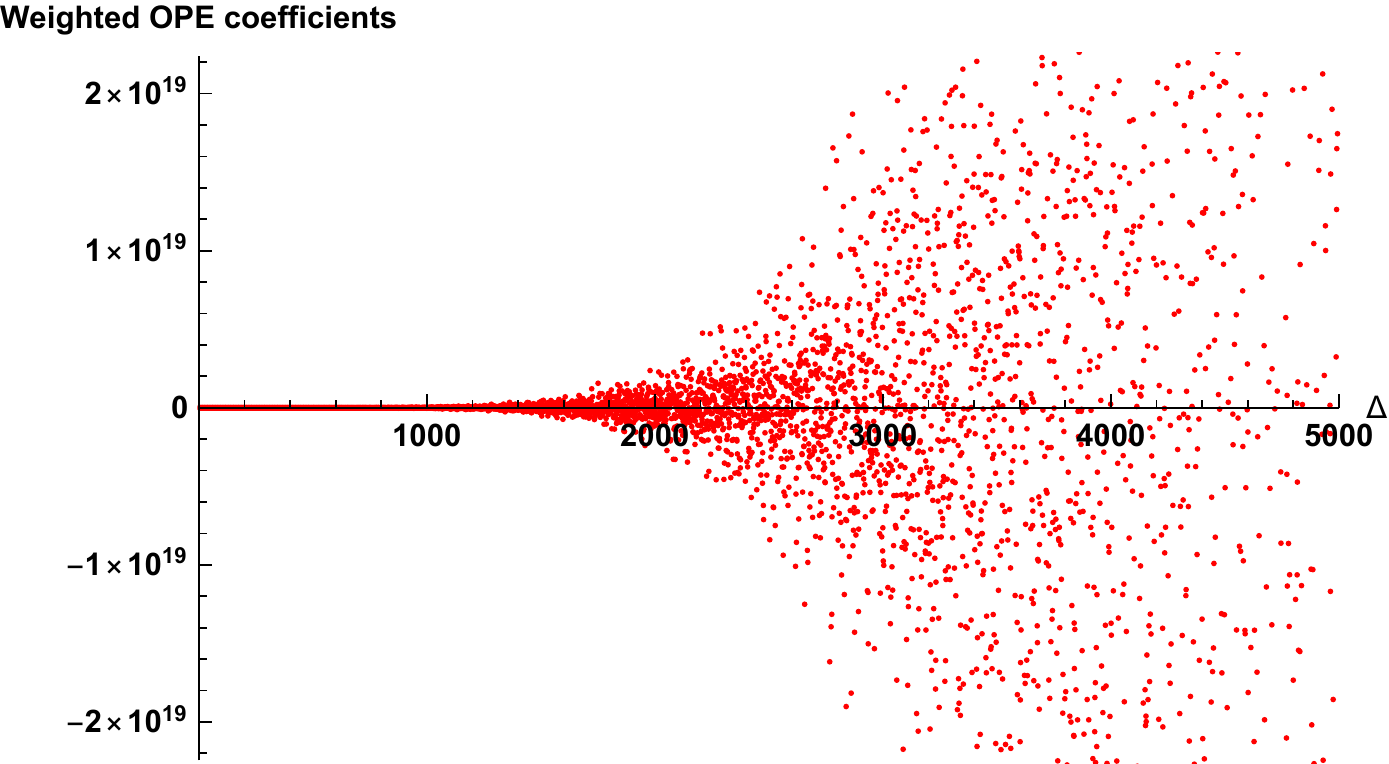}
\includegraphics[scale=0.5]{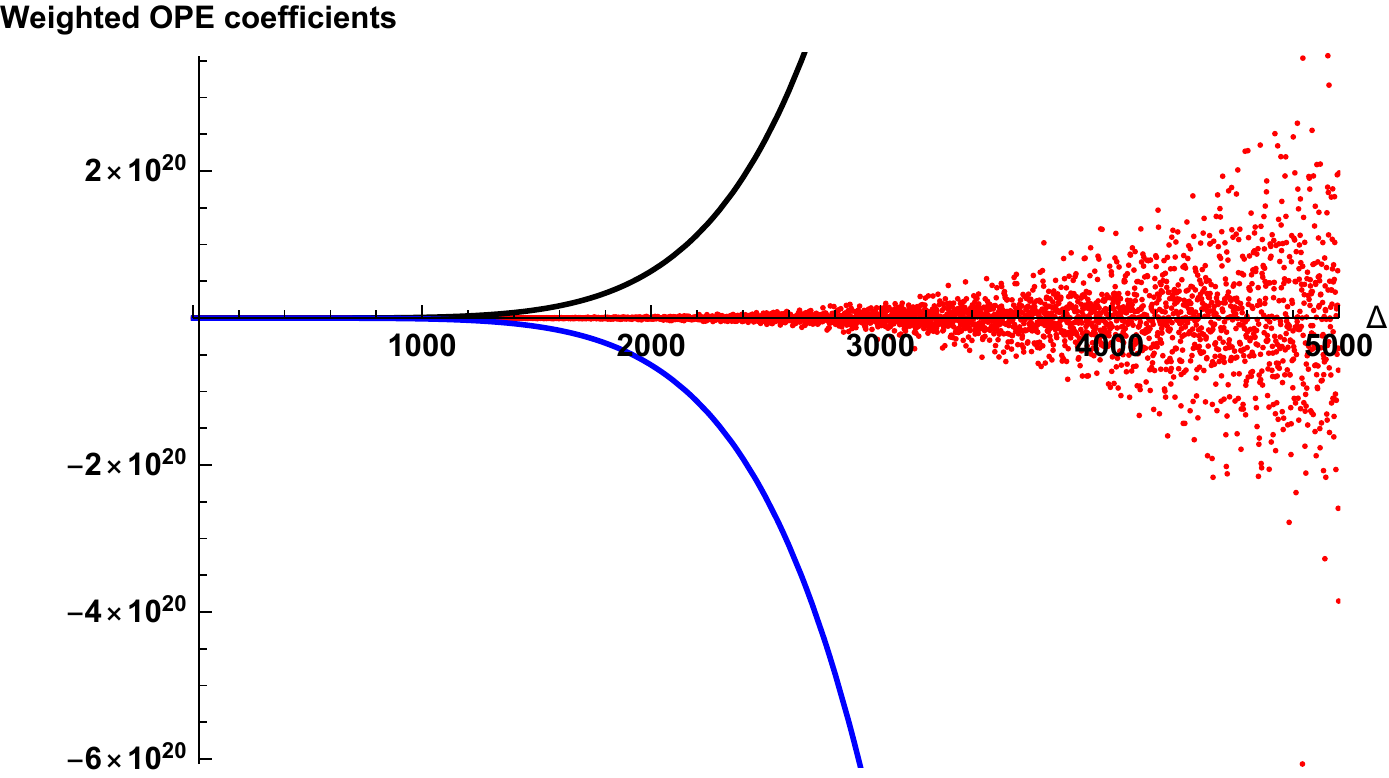}
\caption{$q$ expansion coefficient of the torus one point function of $\otimes^{12}\epsilon=\eta^{24}$ and the fact that $|b_N|$ is bounded by $N^{6}$, denoted by the black and blue line.}
\label{fig:eta24}
\end{figure}

The implication of modular invariance in the  density of high energy states has been analyzed rigorously in \cite{Baur} leveraging the powerful machinery of complex Tauberian theorems. The fact that the Tauberian theorems can be extremely useful in context of CFT is pointed out in \cite{pappadopulo2012operator}, subsequently, the same is emphasized in Appendix C of \cite{dattadaspal}, where the authors have used Ingham's theorem \cite{ingham1941tauberian}. The usefulness of making $\beta$ complex while using Tauberian theorems and thereby gaining extra mileage in controlling the correction terms in various asymptotic quantities of CFT has been pointed out in \cite{mukhametzhanov2018analytic} using the machinery developed in \cite{subhankulov1976tauberian}. One of the pivotal ingredient in the proof and the error estimation in Cardy formula as done in \cite{Baur} is the positive definiteness of the spectral density. Thus the method is not directly adaptable when estimating the asymptotic behavior of three point coefficients \cite{KM}, since three point coefficients can be negative as well (see fig.~\ref{fig:eta} and \ref{fig:eta24}). This has been emphasized repeatedly in \cite{Baur,dattadaspal} and in section 6.3 of \cite{Qiao:2017xif}, where they provided an explicit example to show why the positivity of the spectral density is really important in the context of the Tauberian theorems. While there is an obstruction in directly adapting the method, from a physical ground, we do expect the exponential suppression of average three point coefficients if the three point coefficients are not wildly fluctuating. This asymptotic behavior has implications in context of Eigenstate Thermalization Hypothesis \cite{deutsch1991quantum,srednicki1994chaos,rigol2008thermalization,PhysRevX.8.021026, Lashkari:2016vgj,Basu:2017kzo,Romero-Bermudez:2018dim,Brehm:2018ipf,Hikida:2018khg,Brehm:2019fyy,Dymarsky:2018lhf,Maloney:2018hdg,Maloney:2018yrz,Datta:2019jeo} since the three point coefficients, probed in \cite{KM} are related to expectation value of some operator in a high energy eigenstate of a CFT on a cylinder. Recently, the generalization of KM result as done in \cite{dattadaspal} is utilized in \cite{Alday:2019qrf}. A universal formula for OPE coefficients, $c_{ijk}$ where two of the $i,j,k$ is heavy is on the card as well via use of crossing kernel \cite{Maloney}. To remind the readers, in \cite{KM}, the modular covariance of torus one point function is used to estimate the asymptotic behavior of heavy-light-heavy three point coefficients, they found exponential suppression, which depends on dimension of heavy operator $\Delta$, central charge of the CFT ($c$) and the dimension ($\Delta_\chi$) of the operator, $\chi$ such that  it produces the light operator (with dimension $\Delta_{\mathcal{O}}$) upon doing operator product expansion with itself and it has the least dimension among all such operators producing the light operator upon doing operator product expansion with itself. The result requires $\Delta_\chi<\frac{c}{12}$. For completeness, here is the result (henceforth we will refer to it as ``KM" result and the overline denotes some sense of averaging) from \cite{KM}:
\begin{align}
\overline{f_{\Delta\mathcal{O}\Delta}} \underset{\Delta\to\infty}{\simeq} f_{\chi\mathcal{O}\chi}\left(\Delta-\frac{c}{12}\right)^{\Delta_{\mathcal{O}}/2}\exp\left[-\frac{\pi c}{3}\sqrt{\frac{12\Delta}{c}-1}\left(1-\sqrt{1-\frac{12\Delta_\chi}{c}}\right)\right]
\end{align}
\\

The aim of this paper is to make progress in understanding the asymptotic behavior of the magnitude of three point coefficients and investigating under which conditions one should expect such exponential suppression. In short, our objective is to provide a rigorous underpinning for the behavior of three point coefficients. To motivate further why such characterization is indeed needed from a theoretical standpoint, let us consider two CFTs $\mathcal{C}_1$ and $\mathcal{C}_2$ with central charge $c_1$ and $c_2$ respectively, we tensor them together to construct a CFT $\mathcal{C}_1\otimes \mathcal{C}_2$ with\footnote{The author thanks John McGreevy to point this out in a different context in 2017.}central charge $c_1+c_2$. Now consider an operator $\mathcal{O}\equiv\mathcal{O}_{1}\otimes\mathbb{I}$, naively the average value of the three point coefficient $f^{\mathcal{C}_1\otimes \mathcal{C}_2}_{\Delta\mathcal{O}\Delta}$ as predicted by \cite{KM} would be controlled by the central charge $c_1+c_2$. On the other hand, using $f^{\mathcal{C}_1\otimes\mathcal{C}_2}_{\Delta\mathcal{O}\Delta}=f^{\mathcal{C}_1}_{\Delta\mathcal{O}_1\Delta}$, we see that $f^{\mathcal{C}_1}_{\Delta\mathcal{O}_1\Delta}$ can only possibly depend on central charge $c_1$, not on $c_2$. This example makes it very clear that one needs to be very precise about what is meant by the average value of three point coefficients. We will see that our result actually resolves this paradox. More examples follow to motivate why we might want to take up a rigorous approach in this direction.\\

An example involving tensored copies of CFT can be constructed where the applicability of Kraus-Maloney result\cite{KM} is very subtle. We consider $2$ copies of $2$D Ising model and one copy of a CFT with central charge $\frac{7}{10}$. The tensored CFT, $\mathbb{C}$, has total central charge $c_{eff}=\frac{17}{10}$. Focussing back to the $2$D Ising model, we note that it has following primary operators: $\mathbb{I}$ with dimension $0$, $\sigma$ with dimension $\frac{1}{8}$ and $\epsilon$ with dimension $1$. The unnormalized torus expectation value (henceforth by torus one point function, we will mean unnormalized torus one point function unless otherwise mentioned)  of $\epsilon$ is proportional to square of Dedekind eta ($\eta^2$) function \cite{di1987critical}.  We know that the three point coefficients that contribute to this one point function involves $\sigma$ and its descendants. Upon doing a $q$ expansion of $\eta^2$:
\begin{align}
\langle\epsilon\rangle_{\beta}\propto\eta^2(\beta)=\sum_{n=\Delta-\frac{1}{8}}a_n e^{-\beta\left(n+\frac{1}{8}-\frac{1}{24}\right)}\,,\quad a_n=\underset{\overset{\text{Descendants}}{\text{at $n\ $th level}}}{\sum} f_{\Delta\epsilon\Delta}
\end{align}
one can deduce that the non-zero three point coefficients $f_{\Delta\epsilon\Delta}$ are typically suppressed by $P(n)$, the partition\footnote{Here, in the partition, each integer can occur atmost twice, reflecting the fact the 2D Ising CFT is in fact free fermion, so we have ``fermionic" partitioning instead of usual bosonic one.} of integer $n$, where $n=\Delta-\frac{1}{8}$. The suppression factor is basically the density of descendants of $\sigma$ with dimension $\Delta$. A list plot of $a_n$ looks as in fig.~\ref{fig:eta}.
At this point, one might argue that the lowest operator that contributes to the one point function of $\epsilon$ is $\sigma$, having dimension $\frac{1}{8}$, which is greater than $\frac{c_{\text{Ising}}}{12}=1/24$, thus the result of Kraus-Maloney\cite{KM} is not really applicable\footnote{We recall that the KM result \cite{KM} requires that the lowest operator has dimension $\Delta_\chi<\frac{c}{12}$.}. To cure this, we consider instead the aforementioned tensored CFT $\mathbb{C}$ and the following operator $\epsilon\otimes\mathbb{I}\otimes\mathbb{I}$. The lowest operator that contributes to the torus one point function of $\epsilon\otimes\mathbb{I}\otimes\mathbb{I}$ is $\sigma\otimes\mathbb{I}\otimes\mathbb{I}$, having dimension $\frac{1}{8}$, this is clearly less than $\frac{c_{eff}}{12}$. On the other hand, the three point coefficients of the tensored CFT still boils down to the three point coefficient of Ising Model and thereby suppressed by $P(n)$. We remark that this exponential suppression coming from the factor of $P(n)$ does not involve the effective central charge. Intuitively, the reason behind the departure from KM result  is when we are looking at $f^{\mathbb{C}}_{\Delta\mathcal{O}\Delta}$ of the tensored CFT, we are scanning over all possible operators with dimension $\Delta_{1}\leq \Delta$ such that $f^{\text{Ising}}_{\Delta_1\epsilon\Delta_1}\neq 0$. Thus the window over which ``averaging" has actually been done to obtain what KM predicts contains widely fluctuating numbers, thus the applicability of KM is really very subtle. As mentioned, this indeed provides us with motivation of precisely defining what we are estimating. We will come back to this with a neat resolution at the end of the penultimate section~\S\ref{notkm} with a simplified example consisting of $4$ copies of $2$D Ising model. There we will see that KM is still valid in an appropriately ``integrated" form.\\

Another example which is morally similar to the above would be to consider the $40$ copies of $2$D Ising model (call it CFT $\mathbb{C}_2$) so that $c_{eff}=20$. Now consider the operator
\begin{align}
\mathbb{O}=\left(\otimes^{12}\epsilon\right)\left(\otimes^{28}\mathbb{I}\right)
\end{align}
We remark that even though the discussion that follows are morally same, the reason we include this is the qualitative difference of $q$ expansion coefficients of $\eta^2$ and $\eta^{24}$ (see \ref{fig:eta} and \ref{fig:eta24}). While $\eta^2$ is a \textit{lacunary} function, $\eta^{24}$ is a holomorphic function. Again, the three point coefficient of the tensored CFT depends only on the first $12$ copies of the Ising CFT, as a result we have
\begin{align}
\langle \otimes^{12}\epsilon \rangle_{\beta} = \sum_{N}b_{N}e^{-\beta\left(N+\frac{3}{2}-\frac{1}{2}\right)}\,,\quad b_N=\underset{\overset{\text{Descendants}}{\text{at $N\ $th level}}}{\sum} f_{\Delta\epsilon\Delta}
\end{align}
where $\Delta=N+\frac{3}{2}$. Thus the three point coefficients in this case is typically suppressed by $b_n/\sum_{j}\left(\prod_{i}P(n_i)\right)$ such that $\sum_{i}n_i=N$. On the other hand, we note that
\begin{align}
\langle \otimes^{12}\epsilon \rangle_{\beta} =\eta^{24}
\end{align}
is a modular cusp form of weight $12$ and it can be shown using the properties of holomorphic modular form that $|b_N|=O(N^6)$ as seen in fig.~\ref{fig:eta24}. Thus we have suppression by a factor of $\sum_{j}\left(\prod_{i}P(n_i)\right)$.

In what follows, we will consider a CFT with central charge $c$ and spectrum of operators with dimension $\Delta_i$. To make a precise sense of averaging, we consider an energy window of $\left[\Delta-\delta,\Delta+\delta\right]$ and probe the following quantities: 
\begin{align}
\mathcal{A}&\equiv \left(\frac{G(\Delta+\delta)-G(\Delta-\delta)}{F(\Delta+\delta)-F(\Delta-\delta)}\right)\,,\\
\mathcal{A}^\prime&\equiv \left(\frac{G^\prime(\Delta+\delta)-G^\prime(\Delta-\delta)}{F(\Delta+\delta)-F(\Delta-\delta)}\right)\,,
\end{align}
where $\delta$ is an $O(1)$ number. We can eventually relax this condition to $\delta\simeq\Delta^{\kappa}$ with $\kappa<1/2$. 
We let $\Delta\to\infty$ and we define the functions $G$ and $G^{\prime}$  in the following way:
\begin{align}
G(\Delta)&=\int^{\Delta}_{0} d\Delta^\prime\ \bigg|\sum_{\Delta_{w}=\Delta^\prime}f_{w\mathcal{O}w}\bigg| \left(\sum_{i}\delta(\Delta^\prime-\Delta_i)\right) e^{-\beta\left(\Delta^\prime-\frac{c}{12}\right)}\,,\\
G^\prime(\Delta)&=\int^{\Delta}_{0} d\Delta^\prime\ \left(\sum_{\Delta_{w}=\Delta^\prime}|f_{w\mathcal{O}w}|\right) \left(\sum_{i}\delta(\Delta^\prime-\Delta_i)\right) e^{-\beta\left(\Delta^\prime-\frac{c}{12}\right)}\,,
\end{align}
while the function $F$ is defined as
\begin{align}
F(\Delta)&=\int^{\Delta}_{0} d\Delta^\prime\ \rho(\Delta^{\prime})e^{-\beta\left(\Delta^\prime-\frac{c}{12}\right)}\,,\quad \rho(\Delta^\prime)=\sum_{i}d(\Delta_i)\delta(\Delta^\prime-\Delta_i)\,,
\end{align}
where $\rho(\Delta^{\prime})$ is the density of states, $d(\Delta_i)$ is the degeneracy. We remark that our analysis is sensitive to $q=e^{-\beta}$ expansion coefficient of torus one point function of $\mathcal{O}$ only, to make it apparent, one can also write
\begin{align}
G(\Delta)&=\int^{\Delta}_{0} d\Delta^\prime\ |a(\Delta^\prime)| \left(\sum_{i}\delta(\Delta^\prime-\Delta_i)\right)  e^{-\beta\left(\Delta^\prime-\frac{c}{12}\right)}\,,\\
\langle\mathcal{O}\rangle_{\beta}&=\sum_{i}\ a(\Delta_i) e^{-\beta\left(\Delta_i-\frac{c}{12}\right)}\,, \quad a(\Delta)=\left(\sum_{\Delta_{w}=\Delta}f_{w\mathcal{O}w}\right)\,. 
\end{align} 
Since, $G^\prime(\Delta)\geq G(\Delta)$ ensures that any lower bound on $G(\Delta)$ is also a lower bound for $G^\prime(\Delta)$.\\

The $\mathcal{A}$ can be thought of as the average absolute value of three point coefficients, where averaging is done over the energy window of width $2\delta$, centered at $\Delta$. If we do not wish to average out by the number of states lying in that window, we should instead be considering the following quantities:
\begin{align}
\mathcal{B}=\frac{1}{2\delta} \left[G(\Delta+\delta)-G(\Delta-\delta)\right]\,,\ \mathcal{B}^\prime=\frac{1}{2\delta} \left[G^\prime(\Delta+\delta)-G^\prime(\Delta-\delta)\right]\,.
\end{align}

Furthermore, the expectation of not wildly fluctuating three point coefficients is encapsulated as the following assumption:
\begin{align}\label{basiccondition}
 |f_{\Delta\mathcal{O}\Delta}|\leq M\left(\Delta-\frac{c}{12}\right)^{2k}\,,\quad k\in\mathbb{N}/2\,,\ M\sim O(1)\,,
\end{align}
where $\mathcal{O}$ is a primary with dimension $\Delta_{\mathcal{O}}$ and even spin\footnote{One can relax this condition to $\mathcal{O}$ being $SL(2,R)$ primary (also known as quasi-primary) without much modification. We comment about the odd spin case in the concluding section.} $s$. This is fairly a mild condition, satisfied by the primaries in the aforementioned examples. The condition is also satisfied by the quasi-primaries in the Identity module. Without loss of generality, we further assume\footnote{If this assumption is wrong, we consider the operator $-\mathcal{O}$ and proceed.} that $(-1)^{s/2}f_{\chi\mathcal{O}\chi}>0$, where $\chi$ is an operator with the lowest dimension (say $\Delta_\chi$) such that $ f_{\chi\mathcal{O}\chi}\neq 0$. Like Kraus-Maloney \cite{KM}, we will also require $\Delta_\chi<\frac{c}{12}$. We define a parameter $\gamma$ for future reference:
\begin{align}\label{def:gamma}
0<\gamma\equiv\sqrt{1-\frac{12\Delta_\chi}{c}}\,.
\end{align}
We relax this condition in the penultimate section~\S\ref{notkm} in a restrictive scenario and obtain an estimate for three point coefficients.\\

It so turns out that, when $\delta\simeq O(1)$, the eq.~\eqref{basiccondition} is not enough to prove something useful\footnote{One might hope to improve this because it is not clear whether Eq.~\eqref{modcond} is necessary.}, since this allows for the three point coefficients becoming arbitrarily negative. This motivates us to further assume some bound on how negative it can become. In particular, we will assume that there exists an $\alpha$ such that
\begin{align}\label{modcond}
a(\Delta)>-g(\Delta)\exp\left[2\pi\alpha\sqrt{\frac{c\Delta}{3}}\right]\,, \quad 0<\alpha\leq\gamma
\end{align}
where $a(\Delta)$ is the coefficient of $q^{\Delta-\frac{c}{12}}$ in the $q$ expansion of the torus one point function of $\mathcal{O}$. For a discreet spectrum, one might wonder about the meaning of defining a function $a(\Delta)$ for all $\Delta$, this is done by setting $a(\Delta')=0$ when $\Delta'$ does not appear in the $q$ expansion. Roughly, for large $\Delta$, this is like\footnote{The $a(\Delta)$ captures the sum of all the $f_{w\mathcal{O}w}$ such the operator $w$ has dimension $\Delta$. Strictly speaking, we made a slight abuse of notation while writing $f_{\Delta\mathcal{O}\Delta}$ since there can be multiple operators with dimension $\Delta$ such the three point coefficient is non-zero.} saying, $f_{\Delta\mathcal{O}\Delta}>-g(\Delta)\exp\left[2\pi(\alpha-1)\sqrt{\frac{c\Delta}{3}}\right]$. Here $g(\Delta)$ is a positive polynomial\footnote{For $\Delta<\frac{c}{12}$, one can define $g(\Delta)=H\left(\frac{c}{12}-\Delta\right)^{2\ell}$. In fact, it turns out that we require the \eqref{modcond} to be true only for $\Delta>\Delta_*$, where $\Delta_*$ is an order one number. Thus one can simply as well define, $g(\Delta)=H\Delta^{2\ell}$.}, defined as
\begin{align}\label{def}
g(\Delta)=H\left(\Delta-\frac{c}{12}\right)^{2\ell}\,.
\end{align}\\
When $\delta\simeq \Delta^{\kappa}$ with $\kappa>0$, we can relax the condition given in \eqref{modcond} by setting $\alpha=\gamma$, without imposing any condition on $\ell$. Below, we state our final result in terms of $\delta$, where it is to be understood that $\delta\simeq\Delta^{\kappa}$ with $0\leq \kappa<1/2$ unless otherwise mentioned. The $\kappa=0$ means that we are looking at an $O(1)$ window. We remark at this point that having a condition on how negative $a(\Delta)$ can get does also appear in the context of Tauberian theorems. We will expound on this in appendix~\S\ref{app:A} with a simple toy example with a hope of possible extension of the result that follows.

\paragraph{Results at Finite Central charge-Average Three Point Coefficient:}
Under the assumption as stated in Eq.~\eqref{basiccondition} and Eq.~\eqref{modcond}, we show that typical average value of magnitude of $f_{\Delta\mathcal{O}\Delta}$ in $\Delta\to\infty $ limit is lower bounded by a universal funtion in $\Delta$, which depends only on the central charge, $\Delta_\chi$ and $\Delta_{\mathcal{O}}$.
\begin{align} \label{bound}
\mathcal{A}^\prime \geq \mathcal{A} &\geq a\ \big|\mathcal{C}_{\Delta\mathcal{O}\Delta}\big|\,,\\
\mathcal{B}^\prime \geq \mathcal{B} & \geq b\ \big|T (\Delta)\big|\,,
\end{align}
where $a$ and $b$ are $O(1)$ numbers if $\kappa=0$. In fact, one can borrow the results from \cite{Ganguly:2019ksp} for estimation of $b$. Without detailing much, here we highlight the main features: $b=0.5$ for $\delta \gamma >1$ while for $\delta\gamma\leq 1$, $b$ is less than $0.5$, where $\gamma$ is defined in \eqref{def:gamma}. The estimation of $a$ can be done in a similar way. When $\kappa\neq 0$ (recall, $\delta\sim\Delta^\kappa$, $\kappa<1/2$), there will be an extra suppression coming from the factor $\frac{1}{2\delta}$ in $\mathcal{B}$.\\

Here $T(\Delta)$ is given by
\begin{align}
\label{defT}T(\Delta)&=\frac{1}{\sqrt{2}}N_{\mathcal{O}}f_{\chi\mathcal{O}\chi}\left(\Delta-\frac{c}{12}\right)^{\Delta_{\mathcal{O}}/2-3/4}\exp\left[4\pi\sqrt{\left(\frac{c}{12}-\Delta_\chi\right)\left(\Delta-\frac{c}{12}\right)}\right]\,,
\end{align}
and $\mathcal{C}_{\Delta\mathcal{O}\Delta}$ is given by
\begin{align}
\label{C}\mathcal{C}_{\Delta\mathcal{O}\Delta}&=N^\prime_{\mathcal{O}}f_{\chi\mathcal{O}\chi}\left(\Delta-\frac{c}{12}\right)^{\Delta_{\mathcal{O}}/2}\exp\left[-\frac{\pi c}{3}\sqrt{\frac{12\Delta}{c}-1}\left(1-\sqrt{1-\frac{12\Delta_\chi}{c}}\right)\right]\,,
\end{align}
where $\Delta_\chi$ is the dimension of the lowest lying nontrivial operator $\chi$ with non-zero $f_{\chi\mathcal{O}\chi}$ and $N_{\mathcal{O}}$, $N^\prime_{\mathcal{O}}$ are given by
\begin{align}
N_{\mathcal{O}}&=\left(\frac{c}{12}-\Delta_\chi\right)^{1/4-\Delta_{\mathcal{O}}/2}\,,\\
N^\prime_{\mathcal{O}}&=\left(\frac{c}{12}-\Delta_\chi\right)^{1/4-\Delta_{\mathcal{O}}/2}\left(\frac{12}{c}\right)^{1/4}\,.
\end{align}
This lower bound matches with the result obtained in \cite{KM} by using saddle point approximation and inverse Laplace transformation of the high temperature behavior of torus one point function of the light operator $\mathcal{O}$. We reemphasize that we require $\Delta_\chi <  \frac{c}{12}$ for the proof to go through. This is same as the requirement in \cite{KM}. We have assumed finite $c$ through out the first part of our calculation. Ideally, as \eqref{bound} is asymptotically true one could have substituted $\Delta-c/12\to\Delta$ in eq.~\eqref{defT} and \eqref{C}. The scenario $\Delta_\chi>c/12$ is dealt with in \S\ref{notkm}. \\

One can easily see that how the lower bound resolves the issue with tensored CFT. Considering the single CFT with central charge $c_1$, the lower bound that we obtain (see the first of the eqs.~\eqref{bound}) is greater than the lower bound obtained by considering the tensored CFT with central charge $c_1+c_2$. Thus there is no contradiction/paradoxical situation when considering three point coefficients involving the operator $\mathcal{O}\equiv\mathcal{O}_{1}\otimes\mathbb{I}$. In fact, the discussion at the end of the \S\ref{notkm} gives more insight to this paradox. In particular we show that only in the integrated form, it makes sense to derive asymptotic behavior rather than making a statement about typical behavior of three point coefficients.\\

As mentioned, the above result can be generalized to an operator with even spin, in particular, one can obtain results for the operator $\mathcal{O}=T+\bar T$ and thus verify against the known result for $f_{\Delta\mathcal{O}\Delta}$. We also verify it for the tensored CFTs, as elucidated in \S\ref{ex}. The implications and further possible extensions of the above is discussed in concluding section~\S\ref{conc}. The appendix~\S\ref{app:A} expounds on some idea regarding how to possibly prove an upper bound in a more generic set up, thus exactly deriving the asymptotic behavior. The appendix~\S\ref{app:B} contains one more example verifying the bound.\\

\paragraph{Results under milder assumption:}Technically one can see that the ineq.~\eqref{modcond} is only needed for $\Delta>\Delta_{*}$ for some order one number $\Delta_{*}$. Thus if the condition is violated, it needs to be violated for an infinite number of $\Delta$. Thus one can always form an increasing subsequence $\{\Delta_{k}$\} such that we have
\begin{align}
|a(\Delta_{k})|\underset{k\to\infty}{\geq} u\left(\Delta-\frac{c}{12}\right)^{\frac{\Delta_{\mathcal{O}}}{2}-\frac{3}{4}}\exp\left[2\pi\gamma\sqrt{\frac{c\Delta_k}{3}}\right]
\end{align}  
for some order one number $u$. In that scenario, we have found an increasing subsequence $\Delta_k$ such that for $\delta<\delta_{gap}$ (assuming that there is a minimal gap in the operator spectra given by $2\delta_{gap}$, recently it has been shown that $2\delta_{gap}\leq 1$ \cite{Ganguly:2019ksp,Baur}), the asymptotic behavior ($k\to\infty$) of $\mathcal{B}$ is bounded below by $g(\Delta_k)\exp\left[2\pi\gamma\sqrt{\frac{c\Delta_k}{3}}\right]$ for some polynomial $g(\Delta)$. This also helps us to gain more insight to the condition \eqref{modcond}. The condition is a statement that all the negative $a(\Delta)$ forms a subsequence such that $\{|a^{\text{negative}}(\Delta)|\}$ is  asymptotically upper bounded by $T(\Delta)$.\\

\paragraph{Asymptotics using Ingham's theorem \cite{ingham1941tauberian}:} Instead of \eqref{modcond}, if one assumes following condition (this is asymptotically equivalent to \eqref{modcond}):
\begin{equation}
f_{w\mathcal{O}w} > - \frac{g(\Delta_w)}{\rho_0(\Delta_w)}\exp\left[2\pi\alpha\sqrt{\frac{\Delta_w}{3}}\right]\,,
\end{equation}
where $\rho_0(\Delta)$ is the Cardy formula for the density of states; then using Ingham's theorem \cite{ingham1941tauberian}, it can be shown\footnote{The author thanks Baur Mukhametzhanov for pointing this out.} that 
\begin{equation}
\int_{0}^{\Delta}\text{d}\Delta'\ a(\Delta') \underset{\Delta\to\infty}{\sim} \frac{1}{\sqrt{2}}f_{\chi\mathcal{O}\chi}\Delta^{\frac{\Delta_{\mathcal{O}}}{2}-\frac{1}{4}}\left(\frac{c}{12}-\Delta_\chi\right)^{-\frac{\Delta_{\mathcal{O}}}{2}-\frac{1}{4}}\exp\left[4\pi\sqrt{\left(\frac{c}{12}-\Delta_\chi\right)\Delta}\right]\,.
\end{equation}

\paragraph{Results at large central charge-Average Three Point Coefficient:}
The lower bound can also be obtained for infinite central charge. In this scenario, the lower bound has explicit dependence on the width of energy window. The precise result is obtained by keeping $\Delta/c$ finite while taking $c\to\infty$. In particular, we assume that
\begin{align}
\Delta=c\left(\frac{1}{12}+\epsilon\right)\,, \quad c\to\infty\,, \epsilon \ \text{fixed}.
\end{align}
We further assume that $\Delta_\chi \sim O(c)$, this is consistent with HKS sparseness condition \cite{HKS} while estimating the density of states.
For $\Delta>\frac{c}{6}$ and for $\delta\sim O(1)$, we derive  
\begin{align} \label{bound2}
\mathcal{A}^\prime \geq \mathcal{A} &\geq a \exp\left[-\frac{\pi\delta\sqrt{1-\frac{12\Delta_\chi}{c}}}{\sqrt{3\epsilon}}\right]\ \bigg|\widetilde{\mathcal{C}}_{\Delta\mathcal{O}\Delta}\bigg|\,,\\
\mathcal{B}^\prime \geq \mathcal{B} & \geq b\exp\left[-\frac{\pi\delta\sqrt{1-\frac{12\Delta_\chi}{c}}}{\sqrt{3\epsilon}}\right] \ \bigg|\widetilde{T} (\Delta)\bigg|\,,
\end{align}
where $a,b$ are $O(1)$ numbers. The $\widetilde{T}(\Delta)$ and $\widetilde{\mathcal{C}}_{\Delta\mathcal{O}\Delta}$ are given by
\begin{align}\label{tildeT}
\widetilde{T}(\Delta)&=\frac{1}{\sqrt{2}}N_{\mathcal{O}}f_{\chi\mathcal{O}\chi}\left(c\epsilon\right)^{\Delta_{\mathcal{O}}/2-3/4}\exp\left[4\pi\sqrt{c\epsilon\left(\frac{c}{12}-\Delta_\chi\right)}\right]\,,\\
\label{tildeC}
\widetilde{\mathcal{C}}_{\Delta\mathcal{O}\Delta}&=N^\prime_{\mathcal{O}}f_{\chi\mathcal{O}\chi}\left(c\epsilon\right)^{\Delta_{\mathcal{O}}/2}\exp\left[-2\pi c\sqrt{\frac{\epsilon}{3}}\left(1-\sqrt{1-\frac{12\Delta_\chi}{c}}\right)\right]\,.
\end{align}
We remark that if $\Delta_\chi << c$, in the leading order, one obtains
\begin{align}\label{resulteth}
\widetilde{\mathcal{C}}_{\Delta\mathcal{O}\Delta}&=N^\prime_{\mathcal{O}}f_{\chi\mathcal{O}\chi}\left(c\epsilon\right)^{\Delta_{\mathcal{O}}/2}\exp\left[-2\pi c\sqrt{\frac{\epsilon}{3}}\left(\frac{6\Delta_\chi}{c}\right)\right]\,.
\end{align}
The above expression is same as the one obtained in \cite{KM} in the following limit $\Delta\to\infty$ and then taking $c\to\infty$.\\
 
The above result is consistent with Eigenstate thermalization hypothesis (ETH) in the sense that the ETH predicts the average value of the expectation value of a primary operator in heavy state is same as its thermal expectation value. Since the magnitude of a number is always bigger than the number, we expect $\mathcal{A}$ to be greater than or equal to the thermal expectation value. One can easily verify that the right hand side of Eq.~\eqref{resulteth} is indeed the thermal expectation value for $\beta_{ETH}=\frac{\pi}{\sqrt{3\epsilon}}$. The $\beta_{ETH}$ perfectly matches with the $\beta$ obtained by solving the following equation for $\beta$
\begin{align}
\langle \Delta | H |\Delta\rangle = \langle H \rangle_{\beta}
\end{align}
In short, we show that 
\begin{align}
\mathcal{A} \gtrapprox \langle\mathcal{O}\rangle_{\beta_{ETH}}\,.
\end{align}

The large central charge behavior reveals that the three point function can not fall off exponentially unless $\Delta_\chi$ is of the order of central charge. In particular, this rules out the possibility of suppression of $f_{\Delta T\Delta}$ at large central charge, since in that case $\Delta_{\chi=\mathbb{I}}=0$. Similar behavior is expected for KdV charges as well \cite{Brehm:2019fyy,Dymarsky:2018lhf,Maloney:2018hdg,Maloney:2018yrz}. We remark that the result can easily be generalized for operators $\mathcal{O}$ with even spin. We comment about the odd spin case in the conclusion. Here again, the statement that we made about finding subsequence holds true if the condition in the ineq.~\eqref{modcond} is not satisfied. For the rest of the paper, we assume that the ineq.~\eqref{modcond} is satisfied unless otherwise mentioned. 
\section{Scheme of the proof}
We consider a CFT on a torus, where the spatial cycle is of length $2\pi$ and the thermal cycle is of length $\beta$. We will be looking at the torus one point function of $\mathcal{O}$, a primary operator of the CFT. For simplicity, we consider a spinless operator\footnote{This can easily be generalized for spinful operator and quasi-primaries.}. The main tool we are going to use is modular covariance property of torus one point function of this operator, $\mathcal{O}$ and modular invariance of partition function $Z(\beta)$. In several of steps below, we will be heavily using triangle inequality in the form $|\int f(x) dx| \leq \int |f(x)| dx$ and the inequality $-|r|\leq r\leq |r|$ for any real number $r$. \\

We will do the proof in several steps:
\begin{enumerate}
\item First of all we will assume $f_{\chi\mathcal{O}\chi}>0$ for the first nontrivial operator $\chi$, that produces operator $\mathcal{O}$ upon doing operator product expansion with itself ($\chi\chi\sim \mathbf{I}+\cdots+\mathcal{O}+\cdots$). Under this assumption, we warm up with the restrictive scenario (instead of considering Eq.~\eqref{basiccondition} ) where $f_{\Delta\mathcal{O}\Delta}\in[-M,M]$ for some positive number $M$. From there, we arrive at \eqref{bound}. This is done in \S\ref{M}. We relax this condition afterwards in \S\ref{M2} and allow for polynomial  growth.
\item The large central charge analysis is done in the \S\ref{largec}.
\item In \S\ref{ex}, we sketch out an extension of the lower bound for operators $\mathcal{O}$ with even spin (and not necessarily primary, rather a quasi-primary) and provide explicit verifications of the bound. 
\end{enumerate}

The basic idea is to have an approximation of the weighted three point coefficient $\mathcal{B}$ by a convolution of torus one point function in dual temperature ($\beta^\prime=\frac{4\pi^2}{\beta}$ and we let $\beta\to0$ at the end) and some auxiliary band limited function\footnote{Band limited function means a function whose Fourier transform has a finite support.} . The torus one point function is then separated into a light part and a heavy part. The light part captures the leading behavior of torus expectation value in high temperature limit and yields the approximation of the weighted three point coefficient $\mathcal{B}$. This is done via a suitable choice of $\beta$ as a function of $\Delta$ where we take $\Delta\to\infty$. Once we have a choice of $\beta$, the contribution to the torus one point function coming from the heavy states are shown to be suppressed. One way to show this is to put a bound on the heavy contribution by a part of partition function (we shall name it $Z_H$ in what follows), which gets contribution from the heavy states (or derivatives of $Z_H$). This, in turn, can be estimated to be bounded by some subleading term. We will see, the HKS bound \cite{HKS} (a bound derived by Hartman, Keller and Stoica) and its modified version will play a crucial role in this step. A more detailed semi-technical review of Tauberian techniques can be found in \cite{Pal:2019zzr}.\\  

One more technical remark is in order whose importance will become clear as we go along. We will be heavily using triangle inequality in the form 
\begin{align}
a(\Delta)=\sum_{\Delta_w=\Delta}f_{w\mathcal{O}w}\leq \bigg|\sum_{\Delta_w=\Delta}f_{w\mathcal{O}w}\bigg|=|a(\Delta)|\,.
\end{align} 
Now one would like to integrate both hand side over the interval $[\Delta-\delta,\Delta+\delta]$ to have a lower bound. This can not be done directly, rather one has to introduce a bandlimited function $\phi_-$, which is less than or equal to the indicator function of the mentioned interval: this is required to make sure that the contribution from the heavy states are cut off. So schematically we have something like:
\begin{align}
\nonumber &e^{\beta(\Delta-\delta)}\int_{\Delta-\delta}^{\Delta+\delta}\ d\Delta^\prime\ a(\Delta^\prime)\sigma(\Delta^\prime)\phi_-(\Delta^\prime) \ e^{-\beta\Delta^\prime}\\
\label{scheme}&\leq e^{\beta(\Delta-\delta)}\int_{\Delta-\delta}^{\Delta+\delta}\ d\Delta^\prime\ a(\Delta^\prime)\sigma(\Delta^\prime) \ e^{-\beta\Delta^\prime} \leq 2\delta \mathcal{B}
\end{align}
where we have introduced $\sigma(\Delta^\prime)$ as a shorthand for the following quantity:
\begin{align}\label{defsigma}
\sigma(\Delta^\prime)= \left(\sum_{i}\delta(\Delta^\prime-\Delta_i)\right)
\end{align}
To relate this to the torus one point function, we have to add to the both sides of the inequality~\eqref{scheme} the contribution coming from the states, not in the interval. This schematically looks like as follows: 
\begin{align}
\nonumber &e^{\beta(\Delta-\delta)}\left(\int_{\Delta-\delta}^{\Delta+\delta}\ d\Delta^\prime\ a(\Delta^\prime)\sigma(\Delta^\prime) \ e^{-\beta\Delta^\prime}+\text{stuff not in the interval}\right)\\
\label{scheme2}& \leq 2\delta \mathcal{B}+\underbrace{e^{\beta(\Delta-\delta)}\left(\text{stuff not in the interval} \right)}_{\text{subleading!}}
\end{align}
Once we get the torus one point function, we can use modular covariance to write down an expression of it at high temperature, to be precise we get some convolution of torus one point function and Fourier transform of $\phi_-$. We then use bandlimited nature of $\phi_-$ to cut off the contribution from the heavy states and show that the leading contribution comes from the low lying states only. But our job is not done yet, since the right hand side of the inequality given by \eqref{scheme2} still has those contribution coming from the states, not in the interval. Our final job would then be to estimate this extra contribution and show they are subleading and does not matter in large $\Delta$ limit. For the analysis of Cardy formula, we would not have to do this extra bit since $\phi_-$ is negative outside the interval and density of states is positive, thus the extra bit is by default negative and we can ignore it. But here the negativity of the extra bit is not really guaranteed.\\ 

The readers who want to circumnavigate the technical details for their first read can now skip directly to the \S\ref{ex}.

\section{Derivation of the result: Warm up}\label{M}

This section deals with a restrictive scenario where we assume $|f_{\Delta\mathcal{O}\Delta}| <M$ for some $O(1)$ number $M$. Later we relax this condition and allow for power law growth, which requires more sophistication and a modified version of a bound derived by Hartman, Keller and Stoica in \cite{HKS}, henceforth called as HKS bound.

\subsection{A Lemma}
Let us divide the contribution from light states and heavy states towards the torus one-point function separately: 
\begin{align}
\langle\mathcal{O}\rangle &= \langle\mathcal{O}\rangle^{L}+\langle\mathcal{O}\rangle^{H}\\
 \langle\mathcal{O}\rangle^{L}&= \sum_{\Delta<\Delta_H} f_{\Delta\mathcal{O}\Delta}\ e^{-\beta(\Delta-c/12)}\,,\quad 
  \langle\mathcal{O}\rangle^{H}= \sum_{\Delta\geq \Delta_H} f_{\Delta\mathcal{O}\Delta}\ e^{-\beta(\Delta-c/12)}
\end{align}
and $\Delta_H > \frac{c}{12}$. Here we have done slight abuse of notation: we mean the right hand side of the following by writing the left hand side: 
\begin{align}
\sum_{\Delta} f_{\Delta\mathcal{O}\Delta}\ e^{-\beta(\Delta-c/12)}\mapsto \sum_{\Delta} \left(\sum_{\Delta_w=\Delta}f_{w\mathcal{O}w}\right)\ e^{-\beta(\Delta-c/12)}
\end{align}  The aim is to bound the $\langle\mathcal{O}\rangle^{H}$ by a part of partition function on which one can apply the HKS bound.
Assuming $|f_{\Delta\mathcal{O}\Delta}|<M\sim O(1)$, the lemma states that 
\begin{align}\label{eq:lemma1}
| \langle\mathcal{O}\rangle^{H}| \leq M\ Z_H\left(Re\left[\beta\right]\right) \,, \quad \text{where}\ Z_H(\beta)= \sum_{\Delta\geq \Delta_H} e^{-\beta(\Delta-c/12)}\,.
\end{align}

The proof follows from the following observation:
\begin{align}
 \bigg|\langle\mathcal{O}\rangle^{H}\bigg|&= \bigg|\sum_{\Delta\geq \Delta_H} f_{\Delta\mathcal{O}\Delta}\ e^{-\beta(\Delta-c/12)}\bigg|\leq \sum_{\Delta\geq \Delta_H}| f_{\Delta\mathcal{O}\Delta}| \ e^{-Re(\beta)(\Delta-c/12)} \leq M Z_{H}\left(Re\left[\beta\right]\right)
\end{align}

\begin{ex}[The Lemma]{Bound on one point function}{}
We will be using the following result later:
\begin{align}\label{eq:lemmaapp}
 \bigg|\langle\mathcal{O}\rangle^{H}_{\frac{4\pi^2}{\beta+\imath t}}\bigg|\leq M\ Z_H\left(\frac{4\pi^2\beta}{\beta^2+t^2}\right)\,, \ \text{where}\ \beta,t \in \mathbb{R}
\end{align}

\end{ex}

\subsection{Main Proof}
The main idea of the proof stems by giving a lower bound to the indicator function for an interval by a band limited function. We remind the readers that the role of this function is to facilitate cutting off the contribution coming from the heavy states to torus one point function. In particular, following \cite{Baur}, we consider a function $\phi_{-}(\Delta^\prime)$ such that we have
\begin{align}\label{eq:functionchoice}
\phi_-(\Delta^\prime) \leq \Theta\left(\Delta^\prime \in [\Delta-\delta,\Delta+\delta]\right)\,.
\end{align}
From this one can arrive at
\begin{align}\label{ineq:1}
e^{\beta(\Delta-\delta)}e^{-\beta\Delta^\prime}\phi_-(\Delta^\prime) \leq \Theta\left(\Delta^\prime \in [\Delta-\delta,\Delta+\delta]\right)\,.
\end{align}
At this point, we define the following positive definite measures using the density of states and the weighted density of states (weighted by absolute value of three point coefficients):
\begin{align}
dF(\Delta^\prime)&= \rho(\Delta^\prime) d\Delta^\prime\,,\\
dG(\Delta^\prime)&= \bigg|\sum_{\Delta_w=\Delta^\prime}f_{w\mathcal{O}w}\bigg|\sigma(\Delta^\prime) d\Delta^\prime=|a(\Delta^\prime)|\sigma(\Delta^\prime) d\Delta^\prime\,.
\end{align}
(we recall the definition of $\sigma$ from the eq.~\eqref{defsigma}) such that \eqref{ineq:1} can be integrated against the measure $dG$ to obtain:
\begin{align}
e^{\beta(\Delta-\delta)} \int_{\Delta-\delta}^{\Delta+\delta} dG(\Delta^\prime) e^{-\beta\Delta^\prime}\phi_-(\Delta^\prime)
&\leq \int_{\Delta-\delta}^{\Delta+\delta} dG(\Delta^\prime) 
\end{align}
Since $a(\Delta^\prime) \leq |a(\Delta^\prime)|$ we also have 
\begin{align}
e^{\beta(\Delta-\delta)} \int_{\Delta-\delta}^{\Delta+\delta} d\Delta^\prime\ a(\Delta^\prime)\sigma(\Delta^\prime)\  e^{-\beta\Delta^\prime}\phi_-(\Delta^\prime) \leq e^{\beta(\Delta-\delta)} \int_{\Delta-\delta}^{\Delta+\delta} dG(\Delta^\prime) e^{-\beta\Delta^\prime}\phi_-(\Delta^\prime) 
\end{align}
where we have assumed $\phi_-(\Delta^\prime)\geq 0$, if $\Delta^\prime\in \left[\Delta-\delta,\Delta+\delta\right]$. We remark that this is not in contradiction with Eq.~\eqref{eq:functionchoice}. Rather it puts more constraint on the possible functions that we can choose. Combining the last two inequalities, we have
\begin{align}\label{master}
e^{\beta(\Delta-\delta)} \int_{\Delta-\delta}^{\Delta+\delta} d\Delta^\prime\ a(\Delta^\prime)\sigma(\Delta^\prime)\  e^{-\beta\Delta^\prime}\phi_-(\Delta^\prime)\leq \int_{\Delta-\delta}^{\Delta+\delta} dG(\Delta^\prime)
\end{align}
Now one obtains from the above: 
\begin{align}
\nonumber e^{\beta\Delta_-} \int_0^\infty  d\Delta^\prime\ a(\Delta^\prime)\sigma(\Delta^\prime)\  e^{-\beta\Delta^\prime}\phi_-(\Delta^\prime)&\leq \int_{\Delta_-}^{\Delta_+} dG(\Delta^\prime)\\
\label{master0}&+\underset{\underset{\Delta^\prime \notin \left[\Delta_-,\Delta_+\right]}{\Delta^\prime>0}}{\int} d\Delta^\prime \sigma(\Delta^\prime)a(\Delta^\prime)\  e^{-\beta[\Delta^\prime-\Delta_-]}\phi_-(\Delta^\prime) \,,
\end{align}
where $\Delta_\pm=\Delta\pm\delta$. Had it been the case that $f_{\Delta\mathcal{O}\Delta}$ is positive everywhere, so is $a(\Delta)$; then we would have obtained the following:
\begin{align}\label{master}
e^{\beta(\Delta-\delta)} \int_0^\infty d\Delta^\prime\sigma(\Delta^\prime)\ a(\Delta^\prime)\ \phi_-(\Delta^\prime)  e^{-\beta\Delta^\prime}\leq \int_{\Delta-\delta}^{\Delta+\delta} dG(\Delta^\prime)\,,
\end{align}
since $\phi_-(\Delta^\prime)$ is negative outside the window $\left[\Delta-\delta,\Delta+\delta\right]$. This is precisely where the problem lies. We know that $f_{\Delta\mathcal{O}\Delta}$ need not be positive for all $\Delta$, hence $a(\Delta)$ need not be positive. One can actually see that the positivity is not really a necessary condition, all we require for the Eq.~\eqref{master} to be true at finite $\beta$ is the following condition:
\begin{align}\label{exotic}
\exp\left[\beta\Delta_-\right]\left(\underset{\underset{\Delta^\prime \notin \left[\Delta_-,\Delta_+\right]}{\Delta^\prime>0}}{\int} d\Delta^\prime\sigma(\Delta^\prime)\  a(\Delta^\prime)\phi_-(\Delta^\prime)e^{-\beta\Delta^\prime}\right) <0
\end{align}
Let us denote the region $\Delta^\prime\notin \left(\left[\Delta-\delta,\Delta+\delta\right]\cup (-\infty,0)\right)$ as $\mathcal{S}$. Certainly if all the $f_{\Delta^\prime\mathcal{O}\Delta^\prime}$ is negative definite for $\Delta \in S$, this is not true. But we can relax this condition further more. As we are interested in $\Delta\to\infty$ limit, we are required to show that the second term in the right hand side of the ineq.~\eqref{master0} is bounded above by some term, which is subleading compared to the two other terms in the ineq.~\eqref{master0}. For now, we assume that it can be done and we proceed. We will come back to this at the end of this section.\\

We also define Laplace transform of density of states and weighted density of states via following:
\begin{align}
\mathcal{L}_\rho(\beta)&\equiv \int_0^{\infty} d\Delta\ \rho(\Delta)\exp(-\beta\Delta)=e^{-\beta\frac{c}{12}}Z(\beta)\\
\mathcal{Y}_\rho(\beta)&\equiv \int_0^{\infty} d\Delta\ a(\Delta) \sigma(\Delta)\exp(-\beta\Delta)=e^{-\beta\frac{c}{12}}\langle\mathcal{O}\rangle_{\beta}
\end{align}
where $Z(\beta)$ is the partition function and $\langle\mathcal{O}\rangle_{\beta}$ is the one-point function of the primary operator $\mathcal{O}$. Furthermore, we define Fourier transform of $\phi_-$ via $\phi_{-}(\Delta)=\int_{-\infty}^{\infty}\hat{\phi}_{-}(t)e^{-\imath\Delta t}$ to rewrite the inequality \eqref{master} in the following form:
\begin{align}\label{master2}
e^{\beta(\Delta-\delta)}\int^{\infty}_{-\infty}\ dt\ \hat{\phi}_{-}(t) \mathcal{Y}_{\rho}(\beta+\imath t) \leq  \int_{\Delta-\delta}^{\Delta+\delta} dG(\Delta^\prime) 
\end{align}

Now let us focus on the quantity $\mathcal{Y}_{\rho}(\beta+\imath t)$ and use modular covariance, which states that
\begin{align}
\langle\mathcal{O}\rangle_{-1/\tau}=\tau^h\bar\tau^{\bar h} \langle\mathcal{O}\rangle_{\tau}
\end{align}
where $\tau $ is the modular parameter of the torus\footnote{We remark that we are treating $\tau$ and $\bar{\tau}$ to be independent variables and set them to $\pm\frac{\imath(\beta+\imath t)}{2\pi}$}. Hence we have
\begin{equation}
\begin{aligned}\label{eqY}
\mathcal{Y}_{\rho}(\beta+\imath t) &= e^{-(\beta+\imath t) c/12} \left(\frac{2\pi}{\beta+\imath t}\right)^{\Delta_{\mathcal{O}}}\left(\langle\mathcal{O}\rangle^{L}_{\frac{4\pi^2}{\beta+\imath t}}+\langle\mathcal{O}\rangle^{H}_{\frac{4\pi^2}{\beta+\imath t}}\right)\\
&=\mathcal{Y}_{\rho_*}(\beta+\imath t) +e^{-(\beta+it) c/12} \left(\frac{2\pi}{\beta+\imath t}\right)^{\Delta_{\mathcal{O}}}\langle\mathcal{O}\rangle^{H}_{\frac{4\pi^2}{\beta+\imath t}} \,,
\end{aligned}
\end{equation}
where we have defined 
\begin{align}
\mathcal{Y}_{\rho_*}(\beta)\equiv e^{-\beta c/12} \left(\frac{2\pi}{\beta}\right)^{\Delta_{\mathcal{O}}}\langle\mathcal{O}\rangle^{L}_{4\pi^2/\beta} \,.
\end{align}
We also note that if $\langle\mathcal{O}\rangle^{L}_{4\pi^2/\beta} $ is dominated by the first excited state with dimension $\Delta_\chi$, which is true in $4\pi^2/\beta\to \infty$ limit, we have
\begin{align}\label{KM}
\mathcal{Y}_{\rho_*}(\beta) = \int_0^{\infty} d\Delta\ T(\Delta) e^{-\beta\Delta}\,.
\end{align}
Here $T(\Delta)$ is given as
\begin{align}
T(\Delta)=2 \pi f_{\chi\mathcal{O}\chi}\left(\frac{c}{12}-\Delta_\chi\right)^{\frac{1-\Delta_{\mathcal{O}}}{2}} \left(\Delta-\frac{c}{12}\right)^{\frac{\Delta_{\mathcal{O}}-1}{2}} I_{\Delta_{\mathcal{O}}-1}\left(4 \pi  \sqrt{\left(\Delta-\frac{c}{12}\right)\left(\frac{c}{12}-\Delta_\chi\right)}\right)\,,
\end{align}
which, in $\Delta\to\infty$ limit, goes as: 
\begin{align}\label{KM}
T(\Delta)\sim \frac{1}{\sqrt{2}}N_{\mathcal{O}}f_{\chi\mathcal{O}\chi}\left(\Delta-\frac{c}{12}\right)^{\Delta_{\mathcal{O}}/2-3/4}\exp\left[4\pi\sqrt{\left(\frac{c}{12}-\Delta_\chi\right)\left(\Delta-\frac{c}{12}\right)}\right]\,.
\end{align}
Here $N_{\mathcal{O}}$ is a $\Delta$ independent factor, given by
\begin{align}
N_{\mathcal{O}}=\left(\frac{c}{12}-\Delta_\chi\right)^{1/4-\Delta_{\mathcal{O}}/2}
\end{align}
The expression \eqref{KM} is what is obtained in \cite{KM} by doing naive saddle point approximation.\\

Now we use \eqref{eqY} and the inequality $-|r|\leq r$ for any real number $r$. In particular, we choose 
\begin{align}
\nonumber r=\int^{\infty}_{-\infty}\ dt\ \hat{\phi}_{-}(t) e^{-(\beta+it) c/12} \left(\frac{2\pi}{\beta+\imath t}\right)^{\Delta_{\mathcal{O}}}\langle\mathcal{O}\rangle^{H}_{\frac{4\pi^2}{\beta+\imath t}}\,,
\end{align}
and the reality of $r$ is guaranteed by choosing $\phi_-(\Delta^\prime)$ to be a real valued function. Using $-|r|\leq r$, we rewrite the inequality~\eqref{master2} as
\begin{align}\label{master3}
\nonumber &e^{\beta(\Delta-\delta)}\bigg(\int^{\infty}_{-\infty}\ dt\ \hat{\phi}_{-}(t) \mathcal{Y}_{\rho_*}(\beta+\imath t)-\bigg| \int^{\infty}_{-\infty}\ dt\ \hat{\phi}_{-}(t) e^{-(\beta+it) c/12} \left(\frac{2\pi}{\beta+\imath t}\right)^{\Delta_{\mathcal{O}}}\langle\mathcal{O}\rangle^{H}_{\frac{4\pi^2}{\beta+\imath t}} \bigg|\bigg) \\
& \leq  \int_{\Delta-\delta}^{\Delta+\delta} dG(\Delta^\prime)
\end{align}
At this point we assume that $\hat{\phi}_-$ has bounded support on $\left[-\Lambda_-,\Lambda_-\right]$ so that we have
\begin{equation}\label{master31}
\begin{aligned}
&\bigg| \int^{\infty}_{-\infty}\ dt\ \hat{\phi}_{-}(t) e^{-(\beta+it) c/12} \left(\frac{2\pi}{\beta+\imath t}\right)^{\Delta_{\mathcal{O}}}\langle\mathcal{O}\rangle^{H}_{\frac{4\pi^2}{\beta+\imath t}} \bigg| \\&=\bigg| \int^{\Lambda_-}_{-\Lambda_-}\ dt\ \hat{\phi}_{-}(t) e^{-(\beta+it) c/12} \left(\frac{2\pi}{\beta+\imath t}\right)^{\Delta_{\mathcal{O}}}\langle\mathcal{O}\rangle^{H}_{\frac{4\pi^2}{\beta+\imath t}} \bigg|
\end{aligned}
\end{equation}
The objective of having a bounded support is to cut off the contribution from heavy states in a nice way, as we will see. The bandlimited function has also been used in the analysis of Cardy formula in \cite{Baur,Ganguly:2019ksp,Pal:2019zzr}. Then we move in the absolute value under the integral given in the eq.~\eqref{master31} to obtain:
\begin{eqnarray}
\label{add}&\bigg| \int^{\Lambda_-}_{-\Lambda_-}\ dt\ \hat{\phi}_{-}(t) e^{-(\beta+it) c/12} \left(\frac{2\pi}{\beta+\imath t}\right)^{\Delta_{\mathcal{O}}}\langle\mathcal{O}\rangle^{H}_{\frac{4\pi^2}{\beta+\imath t}} \bigg|\\
\nonumber &\leq  \int^{\Lambda_-}_{-\Lambda_-}\ dt\ |\hat{\phi}_{-}(t)| e^{-\beta c/12} \left(\frac{2\pi}{\sqrt{\beta^2+t^2}}\right)^{\Delta_{\mathcal{O}}}\bigg|\langle\mathcal{O}\rangle^{H}_{\frac{4\pi^2}{\beta+\imath t}} \bigg|\\
\nonumber& 
\leq M\int^{\Lambda_-}_{-\Lambda_-}\ dt\ |\hat{\phi}_{-}(t)| e^{-\beta c/12} \left(\frac{2\pi}{\sqrt{\beta^2+t^2}}\right)^{\Delta_{\mathcal{O}}}Z_{H}\left(\frac{4\pi^2\beta}{\beta^2+t^2}\right)\\
\nonumber&\leq M e^{-\beta c/12} \left(\frac{2\pi}{\beta}\right)^{\Delta_{\mathcal{O}}}Z_{H}\left(\frac{4\pi^2\beta}{\beta^2+\Lambda_-^2}\right)\int^{\Lambda_-}_{-\Lambda_-}\ dt\ |\hat{\phi}_{-}(t)|
\end{eqnarray}
where in the second inequality, we have used \eqref{eq:lemmaapp} and in the last inequality we have used monotonicity of $Z_{H}\left(\frac{4\pi^2\beta}{\beta^2+t^2}\right)$ as a function of $t$ in $\left[-\Lambda_-,\Lambda_-\right]$.

Now we recast \eqref{master3} using the above inequality:
\begin{align}
\nonumber &e^{\beta(\Delta-\delta)}\bigg(\int^{\infty}_{-\infty}\ dt\ \hat{\phi}_{-}(t) \mathcal{Y}_{\rho_*}(\beta+\imath t)-M e^{-\beta c/12} \left(\frac{2\pi}{\beta}\right)^{\Delta_{\mathcal{O}}}Z_{H}\left(\frac{4\pi^2\beta}{\beta^2+\Lambda_-^2}\right)\int^{\Lambda_-}_{-\Lambda_-}\ dt\ |\hat{\phi}_{-}(t)|\bigg) \\
& \leq  \int_{\Delta-\delta}^{\Delta+\delta} dG(\Delta^\prime)
\end{align}

which can then be turned into the following using \eqref{KM}:
\begin{align}\label{master4}
&\nonumber e^{\beta(\Delta-\delta)} \bigg(\int^{\infty}_0 d\Delta^\prime\  T(\Delta^\prime) e^{-\beta\Delta^\prime}\phi_-(\Delta^\prime)-M e^{-\beta c/12} \left(\frac{2\pi}{\beta}\right)^{\Delta_{\mathcal{O}}}Z_{H}\left(\frac{4\pi^2\beta}{\beta^2+\Lambda_-^2}\right)\int^{\Lambda_-}_{-\Lambda_-}\ dt\ |\hat{\phi}_{-}(t)|\bigg) \\
& \leq  \int_{\Delta-\delta}^{\Delta+\delta} dG(\Delta^\prime)
\end{align}
Now our aim is to show that in the large $\Delta$, the second term in the first line is subleading; we will achieve this by making sure that $\Lambda_-$ is less than some threshold value.\\

In order to do that, let us look at the first term. The first term can be evaluated by saddle point method (this discussion is similar to that of Section.~$4.1$ in \cite{Baur}, the only difference is that here we choose $\beta=\pi \gamma\sqrt{\frac{c}{3\Delta}}$ and $\gamma=\sqrt{1-\frac{12\Delta_\chi}{c}}$ to make sure that the saddle is at $\Delta^\prime=\Delta$ \footnote{Ideally, we should have chosen 
\begin{align*}
\beta=\pi \gamma\sqrt{\frac{c}{3\Delta}}\left[1+ \frac{\Delta_{\mathcal{O}}/2-3/4}{2\pi\sqrt{\Delta\left(\frac{c}{12}-\Delta_\chi\right)}}\right]
\end{align*}
We also remark that the value of $\beta$ chosen here matches with the saddle point $\beta_*$ in \cite{KM} only at large $\Delta$, the subleading correction are different, but this is of no consequence.}.) and given by
\begin{align}
e^{\beta(\Delta-\delta)}\int^{\infty}_0 d\Delta^\prime\  T(\Delta^\prime) e^{-\beta\Delta^\prime}\phi_-(\Delta^\prime)=2\delta c_-T(\Delta)\,, \quad \text{where}\ c_-=\frac{1}{2}\int^{\infty}_{-\infty}dx\ \phi_-(\Delta+\delta x) 
\end{align}
where we have assumed that $\phi_-$ is chosen in a way $c_-$ exists and $2\delta c_-$ is an $O(1)$ number. One can always achieve that by choosing an even integer $\nu$ such that $\nu>2$ (so that the function goes to zero as $\Delta^\prime\to \infty$) and $\phi_-$ is given by (more detailed choice of useful functions can be found in \cite{Baur,Ganguly:2019ksp}):
\begin{align}
\phi_{-}(\Delta^\prime)=\left(\frac{\sin\left(\frac{\Lambda_-(\Delta^\prime-\Delta)}{\nu}\right)}{\frac{\Lambda_-(\Delta^\prime-\Delta)}{\nu}}\right)^{\nu}\left(1-\left(\frac{\Delta^\prime-\Delta}{\delta}\right)^2\right)
\end{align}

The second term in the first line of \eqref{master4} is already estimated in \cite{Baur}, but unlike them, here we are choosing $\beta=\pi \gamma\sqrt{\frac{c}{3\Delta}}$. So we can not directly use their estimate. Upon restimating the second piece, we obtain:
\begin{align}
e^{\beta\Delta}Z_{H}\left(\frac{4\pi^2\beta}{\beta^2+\Lambda_-^2}\right)\sim\rho_0(\Delta)^{\gamma+\frac{1}{2\gamma}\left(\frac{\Lambda_-^2}{4\pi^2}-\gamma^2\right)}
\end{align}
Hence the second piece goes like 
\begin{align}\label{first}
M  \left(\frac{2\pi}{\gamma\pi\sqrt{\frac{c}{3\Delta}}}\right)^{\Delta_{\mathcal{O}}} \rho_0(\Delta)^{\gamma+\frac{1}{2\gamma}\left(\frac{\Lambda_-^2}{4\pi^2}-\gamma^2\right)}\int^{\Lambda_-}_{-\Lambda_-}\ dt\ |\hat{\phi}_{-}(t)|
\end{align}
Thus the second term is subleading in large $\Delta$ limit if the exponential growth in $T(\Delta)$ is bigger than the growth of $\rho_0(\Delta)^{\gamma+\frac{1}{2\gamma}\left(\frac{\Lambda_-^2}{4\pi^2}-\gamma^2\right)}$, this boils down to some upper bound on $\Lambda_-$, which depends on $\Delta_\chi$, to be specific we have
\begin{align}\label{cond}
\left(\frac{\Lambda_-}{2\pi}\right)< \gamma=\sqrt{1-\frac{12\Delta_\chi}{c}}\,.
\end{align}
The existence of $\Lambda_-$ requires $\Delta_\chi<\frac{c}{12}$. In \cite{Baur} for the analysis of density of states, $\Delta_\chi$ is effectively $0$, hence $\Lambda_-=2\pi$. Hence if choose $\phi_-$ in a way such that this condition \eqref{cond} is satisfied, we have
\begin{align}
c_-T(\Delta) \leq \frac{G(\Delta+\delta)-G(\Delta-\delta)}{2\delta}\,.
\end{align}
On the other hand we also know that \cite{Baur}:
\begin{align}
c^\prime_-\rho_0(\Delta) \leq \frac{F(\Delta+\delta)-F(\Delta-\delta)}{2\delta} \leq  c^\prime_+ \rho_0(\Delta)\,,
\end{align}
where we have used $c^{\prime}_{\pm}$ to signify that one can in principle choose different function for estimating the density of states. In leading order $\rho_0(\Delta)$ is given by
\begin{align}
\rho_0(\Delta)\underset{\Delta\to\infty}{=} \left(\frac{c}{48\Delta^3}\right)^{1/4}\exp\left(2\pi\sqrt{\frac{c\Delta}{3}}\right)\,.
\end{align} 
Thus combining everything, we have
\begin{align}
\frac{c_-}{c^\prime_+} \mathcal{C}_{\Delta\mathcal{O}\Delta}
\leq \frac{G(\Delta+\delta)-G(\Delta-\delta)}{F(\Delta+\delta)-F(\Delta-\delta)} \leq M\,,\quad \text{where}\ \mathcal{C}_{\Delta\mathcal{O}\Delta}= \frac{T(\Delta)}{\rho_0(\Delta)}
\end{align} 
As a last step, one can basically optimize the ratio $\frac{c_-}{c^\prime_+} $ by choosing different functions, but they will depend on lowest excited state $\Delta_\chi$. Using the functions in \cite{Ganguly:2019ksp}, one can show that the best achievable value of $c_-=0.5$. The only difference from \cite{Ganguly:2019ksp} is that while one can achieve $c_-=0.5$ for $\delta>1$ for the analysis of density of states, here it can only be achieved for $\delta>\frac{1}{\gamma}\geq 1$. A more refined analysis along the lines of \cite{Ganguly:2019ksp} can be undertaken as well, the bounds would remain same, the validity regimes will be scaled by a factor of $\frac{1}{\gamma}$. We also remark that the upper bound is trivial and follows directly from the assumption that $|f_{\Delta\mathcal{O}\Delta}|<M$.

Now, as promised, we come back to the task of showing that the third piece in the eq.~\eqref{master0} is in fact subleading. For $\delta\simeq O(1)$, we have to bring in the extra condition, given in the eq.~\eqref{modcond} on how negative the three point coefficient can get:
\begin{align}
a(\Delta) > -g(\Delta) \exp\left[2\pi\alpha\sqrt{\frac{c\Delta}{3}}\right]
\end{align}
This implies that 
\begin{equation}\label{int}
\begin{aligned}
I&=\exp\left[\beta\Delta_-\right]\left(\underset{\underset{\Delta^\prime \notin \left[\Delta-\delta,\Delta+\delta\right]}{\Delta^\prime>0}}{\int} d\Delta^\prime\  e^{-\beta\Delta^\prime}a(\Delta^\prime)\sigma(\Delta^\prime)\phi_-(\Delta^\prime)\right) \\
&\underset{\sim}{<}\exp\left[\beta\Delta_-\right]\left(\underset{\underset{\Delta^\prime \notin \left[\Delta_-,\Delta_+\right]}{\Delta^\prime>0}}{\int} d\Delta^\prime\  e^{-\beta\Delta^\prime}g(\Delta^\prime)\exp\left[2\pi\alpha\sqrt{\frac{c\Delta^\prime}{3}}\right]|\phi_-(\Delta^\prime)|\right)\\
&<\exp\left[\beta\Delta_-\right] \int_{0}^{\infty}d\Delta^\prime\  e^{-\beta\Delta^\prime}g(\Delta^\prime)\exp\left[2\pi\alpha\sqrt{\frac{c\Delta}{3}}\right] |\phi_-(\Delta^\prime)|\\
&<kg(\Delta)
\exp\left(\pi\left(\gamma+\frac{\alpha^2}{\gamma}\right)\sqrt{\frac{c\Delta}{3}}\right) < T(\Delta) \simeq \Delta^{\Delta_{\mathcal{O}}/2-3/4}\exp\left(2\pi\gamma\sqrt{\frac{c\Delta}{3}}\right) 
\end{aligned}
\end{equation}
where $k$ is an order one number, we have used $\beta=\pi\gamma\sqrt{\frac{c}{3\Delta}}$ followed by a clever use of saddle point approximation and the constraint $\alpha<\gamma$. If $\alpha=\gamma$, then the suppression is by a polynomial piece, for which we needed the extra constraint on $\ell$.  We remark that to derive the above, we use the fact that as $\beta \to 0$, the gap in the spectra remains order one and it does not scale. This is required for going from the first line to second line of the ineq.~\eqref{int}. We have ignored this order one number, hence put a $\sim$ symbol in the inequality. Even without appealing to this argument involving gap, we can be more rigorous in justifying the inequalities, leading to subleading nature of $I$ by doing the assuming the following instead of eq.~\eqref{modcond} (they are same condition asymptotically):
\begin{equation}\label{conditionnew}
f_{w\mathcal{O}w} > - r(\Delta_w)
\end{equation} 
where the operator $w$ has dimension $\Delta_w$ and $r(\Delta')=\frac{g(\Delta')}{\rho_0(\Delta')}\exp\left[2\pi\alpha\sqrt{\frac{c\Delta^\prime}{3}}\right]$, $\rho_0(\Delta')$ being the Cardy formula for density of states. Now we have
\begin{equation}
\begin{aligned}
I&<\exp\left[\beta\Delta_-\right]\int_{\Delta'>0} dF(\Delta^\prime)\  r(\Delta^\prime) e^{-\beta\Delta^\prime}|\phi_-(\Delta')|\\
\end{aligned}
\end{equation}
 Now we note that 
$$\int_{0}^{\infty}dF(\Delta')\ r(\Delta')e^{-\beta\Delta'}|\phi_-(\Delta')|=\int_{0}^{\infty}\left(\beta r(\Delta')-\frac{dr(\Delta')}{d\Delta'}\right)F(\Delta')e^{-\beta\Delta'}|\phi_-(\Delta')|\,,$$ which leads to
$$e^{\beta\Delta}\int_{0}^{\infty}dF(\Delta')\ r(\Delta')e^{-\beta\Delta'} |\phi_-(\Delta')|\underset{\Delta\to\infty}{=}g(\Delta) \exp\left[2\pi\alpha\sqrt{\frac{c\Delta}{3}}\right]\,.$$ Here we have used $-\frac{dr(\Delta')}{d\Delta'}$ has same exponential growth as $r(\Delta')$ for large $\Delta'$ and $F(\Delta')$ has exponential growth followed by a saddle point approximation. Now one runs the similar argument involving constraint on $\alpha$. Intuitively this alternative argument tells us that adding $r(\Delta_w)$ to $f_{w\mathcal{O}w}$ does not spoil the leading high temperature behavior and restores positivity due to \eqref{conditionnew}. In fact this sort of argument (also see \S\ref{app:A}) coupled with Ingham's theorem \cite{ingham1941tauberian} can be used (one can apply Ingham's theorem directly to $r(\Delta_w)+f_{w\mathcal{O}w}$) to deduce an estimate of $a(\Delta)$ as $\Delta\to\infty$, without proving any upper or lower bound\footnote{The author thanks Baur Mukhametzhanov for discussion along this line and pointing this out.}.\\

The case where $\delta\simeq \Delta^\kappa$ with $0<\kappa<1/2$ involves an argument more or less similar to the previous one. In particular, we have
\begin{align}
I&\leq\exp\left[\beta(\Delta-\delta)\right]\left(\underset{\underset{\Delta^\prime \notin \left[\Delta-\delta,\Delta+\delta\right]}{\Delta^\prime>0}}{\int} d\Delta^\prime\  e^{-\beta\Delta^\prime}g(\Delta^\prime)\exp\left[2\pi\gamma\sqrt{\frac{c\Delta}{3}}\right]|\phi_-(\Delta^\prime)|\right)
\end{align}
\begin{align}
&<|\Delta^{-\kappa}|^{\nu}\exp\left(2\pi\gamma\sqrt{\frac{c\Delta}{3}}\right) < T(\Delta) 
\end{align}
We have now taken $\alpha=\gamma$ and the suppression comes from $\phi_-(\Delta^\prime)$ evaluated at $\Delta'=(\Delta\pm\Delta^\kappa)$ leading to $|\Delta^{-\kappa}|^{\nu}$, where we have used the fact that $\kappa\nu>0$. The $\alpha<\gamma$ case can be dealt with using the previous estimation.

\section{Allowing for Power Law growth}\label{M2}

In this section, we relax the condition on $f_{\Delta\mathcal{O}\Delta}$ to the following: 
\begin{align}
|f_{\Delta\mathcal{O}\Delta}| < M\left(\Delta-\frac{c}{12}\right)^{2k}\,, \quad \Delta>\frac{c}{12}\,,\ M\sim O(1)\,;\ k\in \mathbb{N}\,.
\end{align}
This is in accordance with eq.~\eqref{basiccondition}, note if the $k$ in the eq.~\eqref{basiccondition} is half integer, we can always round it upto the next integer bigger than that. Thus without loss of generality we assume $k\in\mathbb{N}$ in this section; this simplifies the calculation as well. The rational behind allowing power law growth is to include the quasi-primaries in the Identity module under the umbrella of our result. For example, for stress-energy tensor $T$, we have $f_{\Delta T\Delta}=(h-c/12)$ and $h+\bar h=\Delta$.\\ 


Here again we are required to show that the contribution coming from the heavy states to the torus one point function is suppressed. The trick is to show that this is bounded above by derivative of $Z_H$. In order to estimate derivative of $Z_H$, we need to have a modified version of the HKS bound \cite{HKS} involving derivatives of partition function with respect to $\beta$. The following subsection achieves this. Similar discussion regarding light state dominance of derivative of partition function can be found in section 5.1 of \cite{Kraus:2018pax}.

\subsection{Modified HKS bounds}
Following the derivation of the usual HKS bound \cite{HKS}, we separate the contribution from the light and the heavy states towards partition function $Z(\beta)$ as 
\begin{align}
Z(\beta)&=Z_L(\beta)+Z_H(\beta)\\
Z_L(\beta)&=\sum_{\Delta<\Delta_H}e^{-\beta(\Delta-c/12)}\,,\quad Z_H(\beta)=\sum_{\Delta>\Delta_H}e^{-\beta(\Delta-c/12)}
\end{align}
where $\Delta_H>\frac{c}{12}$. Modular invariance implies that 
\begin{align}
Z_L+Z_H&=Z^\prime_L+Z^\prime_H\\
\Rightarrow \left(\frac{\partial}{\partial\beta}\right)^{2k}Z_L+\left(\frac{\partial}{\partial\beta}\right)^{2k}Z_H&=\left(\frac{\partial}{\partial\beta}\right)^{2k}Z^\prime_L+\left(\frac{\partial}{\partial\beta}\right)^{2k}Z^\prime_H
\end{align}
where prime denotes that the partition function is evaluated at $\beta^\prime=\frac{4\pi^2}{\beta}$ i.e $Z_{L/H}^{\prime}(\beta)=Z_{L/H}(\beta^{\prime})$. For notational simplicity let us define
\begin{align}
Z_{Lk}= \left(\frac{\partial}{\partial\beta}\right)^{2k}Z_L\,,&\quad Z_{Hk}= \left(\frac{\partial}{\partial\beta}\right)^{2k}Z_H\\
Z^\prime_{Lk}= \left(\frac{\partial}{\partial\beta^\prime}\right)^{2k}Z^\prime_L\,,&\quad Z^\prime_{Hk}= \left(\frac{\partial}{\partial\beta^\prime}\right)^{2k}Z^\prime_H
\end{align}

We note that if $\beta>2\pi$, we have 
\begin{align}
\nonumber Z_{Hk}&=\sum_{\Delta>\Delta_H} \left(\Delta-\frac{c}{12}\right)^{2k}e^{-(\beta-\beta^\prime)(\Delta-\frac{c}{12})}e^{-\beta^\prime(\Delta-\frac{c}{12})} \leq e^{-(\beta-\beta^\prime)(\Delta_H-\frac{c}{12})} \left(\frac{\beta^2}{4\pi^2}\frac{\partial}{\partial\beta}\right)^{2k} Z^\prime_{H} \\
&=e^{-(\beta-\beta^\prime)(\Delta_H-\frac{c}{12})} \left(\frac{\beta^2}{4\pi^2}\frac{\partial}{\partial\beta}\right)^{2k} \left(Z_{H}+Z_{L}-Z^\prime_{L}\right)
\end{align}
Upon using the fact that $\Delta_H>\frac{c}{12}$, we derive in $\beta\to\infty$ limit\footnote{We choose $\beta$ large enough so that $\left(\frac{\beta^2}{4\pi^2}\right)^{2k} e^{-(\beta-\beta^\prime)(\Delta_H-\frac{c}{12})}<1$.}:
\begin{align}\label{caut1}
Z_{Hk} \leq \frac{ \left(\frac{\beta^2}{4\pi^2}\right)^{2k} e^{-(\beta-\beta^\prime)(\Delta_H-\frac{c}{12})} }{1-\left(\frac{\beta^2}{4\pi^2}\right)^{2k} e^{-(\beta-\beta^\prime)(\Delta_H-\frac{c}{12})} }\left(Z_{Lk}-\left(\frac{\beta^2}{4\pi^2}\right)^{-2k} Z^\prime_{Lk}\right)
\end{align}

Modular invariance tells us that in the $\beta\to\infty$ limit, we have
\begin{align}\label{caut2}
\left(\frac{\beta^2}{4\pi^2}\right)^{-2k} Z^{\prime}_{Hk}=-\left(\frac{\beta^2}{4\pi^2}\right)^{-2k} Z^\prime_{Lk}+Z_{Lk}+Z_{Hk}
\end{align}
Thus we arrive at
\begin{align}
Z^{\prime}_{Hk} \leq \frac{\left(\left(\frac{\beta^2}{4\pi^2}\right)^{2k} Z_{Lk}-Z^\prime_{Lk}\right)}{1-\left(\frac{\beta^2}{4\pi^2}\right)^{2k} e^{-(\beta-\beta^\prime)(\Delta_H-\frac{c}{12})} }\,,\quad  \beta \to\infty
\end{align}
Exchanging $\beta$ and $\beta^\prime$ we obtain
\begin{align}
Z_{Hk} \leq  \frac{\left(\left(\frac{4\pi^2}{\beta^2}\right)^{2k} Z^\prime_{Lk}-Z_{Lk}\right)}{1-\left(\frac{4\pi^2}{\beta^2}\right)^{2k} e^{-(\beta^\prime-\beta)(\Delta_H-\frac{c}{12})} }\,,\quad  \beta \to0
\end{align}
In particular, we will be needing the following estimate:
\begin{align}\label{hksm}
Z_{Hk} \simeq \left(\frac{4\pi^2}{\beta^2}\right)^{2k} Z^\prime_{Lk}\,,\quad \beta\to0
\end{align}

\subsection{Estimation of three point coefficient}
Instead of repeating all the basic details, we start with the inequality as given in the ineq.~\eqref{add} and do the estimation in the following manner:
\begin{eqnarray}
\nonumber&\bigg| \int^{\Lambda_-}_{-\Lambda_-}\ dt\ \hat{\phi}_{-}(t) e^{-(\beta+it) c/12} \left(\frac{2\pi}{\beta+\imath t}\right)^{\Delta_{\mathcal{O}}}\langle\mathcal{O}\rangle^{H}_{\frac{4\pi^2}{\beta+\imath t}} \bigg|\\
\nonumber&\leq  \int^{\Lambda_-}_{-\Lambda_-}\ dt\ |\hat{\phi}_{-}(t)| e^{-\beta c/12} \left(\frac{2\pi}{\sqrt{\beta^2+t^2}}\right)^{\Delta_{\mathcal{O}}}\bigg|\langle\mathcal{O}\rangle^{H}_{\frac{4\pi^2}{\beta+\imath t}} \bigg|\\
\nonumber& 
\leq M\int^{\Lambda_-}_{-\Lambda_-}\ dt\ |\hat{\phi}_{-}(t)| e^{-\beta c/12} \left(\frac{2\pi}{\sqrt{\beta^2+t^2}}\right)^{\Delta_{\mathcal{O}}} \left[\left(\frac{\partial}{\partial\beta}\right)^{2k}Z_{H}\right]\left(\frac{4\pi^2\beta}{\beta^2+t^2}\right)\\
&\leq e^{-\beta c/12}M \left(\frac{2\pi}{\beta}\right)^{\Delta_{\mathcal{O}}} Z_{Hk}\left(\frac{4\pi^2\beta}{\beta^2+\Lambda_*^2}\right)\int^{\Lambda_-}_{-\Lambda_-}\ dt\ |\hat{\phi}_{-}(t)|
\end{eqnarray}
where we have used
\begin{equation}
\begin{aligned}
\frac{1}{M}\bigg|\langle\mathcal{O}\rangle^{H}_{\frac{4\pi^2}{\beta+\imath t}} \bigg| &\leq \sum_{\Delta>\Delta_H}\frac{|f_{\Delta\mathcal{O}\Delta}|}{M}e^{-\frac{4\pi^2\beta}{\beta^2+t^2}}\\
& \leq \sum_{\Delta>\Delta_H}\left(\Delta-\frac{c}{12}\right)^{2k}e^{-\frac{4\pi^2\beta}{\beta^2+t^2}}= \left[\left(\frac{\partial}{\partial\beta}\right)^{2k}Z_{H}\right]\left(\frac{4\pi^2\beta}{\beta^2+t^2}\right)
\end{aligned}
\end{equation}
and the fact that $\left[\left(\frac{\partial}{\partial\beta}\right)^{2k}Z_{H}\right]\left(\frac{4\pi^2\beta}{\beta^2+t^2}\right)$ attains a maximum at $t=\Lambda_-$. Again, we need to show that the above piece is subleading compared to the first piece in the ineq.~\eqref{master3}. We do that by estimating $Z_{Hk}\left(\frac{4\pi^2\beta}{\beta^2+\Lambda_-^2}\right)$ via \eqref{hksm} and choosing $\beta=\pi\gamma\sqrt{\frac{c}{3\Delta}}\to 0$ (this is similar to the step done in \cite{Baur} while estimating the contribution to the partition function from heavy states): 
\begin{align}
\nonumber Z_{Hk}\left(\frac{4\pi^2\beta}{\beta^2+\Lambda_-^2}\right) &\simeq \left(\frac{2\pi\beta}{\left(\beta^2+\Lambda_-^2\right)}\right) ^{-4k}Z^{\prime}_{Lk}\left(\frac{4\pi^2\beta}{\beta^2+\Lambda_-^2}\right) = \left(\frac{4\pi \beta}{\sqrt{\frac{c}{3}}(\beta^2+\Lambda_-^2)}\right) ^{-4k}e^{\frac{c(\beta^2+\Lambda_-^2)}{12\beta}}\\
&\simeq  \left(\frac{\Lambda_-^2\sqrt{\Delta}}{4\pi^2\gamma}\right)^{4k}\exp\left[{\frac{\Lambda_-^2}{4\pi^2}\frac{\pi\sqrt{\frac{c\Delta}{3}}}{\gamma}}\right]
\end{align}

Hence the second piece goes like 
\begin{align}
\left(\frac{2}{\gamma\sqrt{\frac{c}{3\Delta}}}\right)^{\Delta_{\mathcal{O}}}   \left(\frac{\Lambda_-^2\sqrt{\Delta}}{4\pi^2\gamma}\right)^{4k}\rho_0(\Delta)^{\gamma+\frac{1}{2\gamma}\left(\frac{\Lambda_-^2}{4\pi^2}-\gamma^2\right)}\int^{\Lambda_-}_{-\Lambda_-}\ dt\ |\hat{\phi}_{-}(t)|
\end{align}
This differs from the eq.~\eqref{first} by a polynomial piece. Hence the rest of the analysis from the earlier section goes through in a straightforward manner. 


\section{Large central charge}\label{largec}
In this section, we will be looking at states with dimension:
\begin{align}
\Delta=c\left(\frac{1}{12}+\epsilon\right)\,,\quad \epsilon>0\ \ \text{fixed}\,, \quad c\to\infty
\end{align}
In this scenario, the asymptotic behavior of the quantity $\mathcal{B}$ is given by Eq.~\eqref{tildeT}. If we divide by the density of states at large central charge, the result is given by Eq.~\eqref{tildeC}. The basic scheme of the proof stays more or less same. Below, we sketch out the salient points for the large central charge analysis.\\

\paragraph{HKS Bound:}
The large central charge analysis requires a careful reconsideration of the modified HKS bound. Earlier, one of the crucial input has been the following approximation:
\begin{align}
\left(\frac{\beta^2}{4\pi^2}\frac{\partial}{\partial\beta}\right)^{2k} Z_{H}&\underset{\beta\to\infty}{=} \left(\frac{\beta^2}{4\pi^2}\right)^{2k} Z_{Hk}\,,\quad \ \left(\frac{\beta^2}{4\pi^2}\frac{\partial}{\partial\beta}\right)^{2k} Z_{L}&\underset{\beta\to\infty}{=} \left(\frac{\beta^2}{4\pi^2}\right)^{2k} Z_{Lk}
\end{align}

This approximation remains true even in the finite $\beta$ but $c\to\infty$ limit as we get the maximum power of $c$ only when all the derivatives act on $Z_{H}$ or $Z_L$. Thus we have
\begin{align}
\left(\frac{\beta^2}{4\pi^2}\frac{\partial}{\partial\beta}\right)^{2k} Z_{H}&\underset{c\to\infty}{=} \left(\frac{\beta^2}{4\pi^2}\right)^{2k} Z_{Hk}\,,\quad \left(\frac{\beta^2}{4\pi^2}\frac{\partial}{\partial\beta}\right)^{2k} Z_{L}&\underset{c\to\infty}{=} \left(\frac{\beta^2}{4\pi^2}\right)^{2k} Z_{Lk}
\end{align} 
The rest of the calculation proceeds in a similar manner and we obtain
\begin{align}
Z_{Hk} \leq  \frac{\left(\left(\frac{4\pi^2}{\beta^2}\right)^{2k} Z^\prime_{Lk}-Z_{Lk}\right)}{1-\left(\frac{4\pi^2}{\beta^2}\right)^{2k} e^{-(\beta^\prime-\beta)(\Delta_H-\frac{c}{12})} }\,,\quad  \beta <2\pi\,,\ c\to\infty
\end{align}
In particular, we have 
\begin{align}
Z_{Hk} \underset{c\to\infty}{\simeq} \left(\frac{4\pi^2}{\beta^2}\right)^{2k}Z^\prime_{Lk}\,,\quad \beta<2\pi
\end{align}
and the right hand side of the above is same as in Eq.~\eqref{hksm}.

\paragraph{Sparseness Condition:}
The second crucial assumption that we have to make here is $\Delta_\chi\sim O(c)$. One can relax this assumption but then there would not be any exponential suppression in the central charge. For example, if one tensors really large number of copies of $2$D Ising model, we have a large central charge, but $\Delta_\chi$ would be still be an order one number. Furthermore, we have to assume that for $\beta>2\pi$, the following sparseness condition holds:
\begin{align}\label{sparseness}
\log\left(1+\sum_{c/12>\Delta>\Delta_\chi}\frac{f_{\Delta\mathcal{O}\Delta}}{f_{\chi\mathcal{O}\chi}}\exp\left[-\beta(\Delta-\Delta_\chi)\right]\right)\underset{c\to\infty}{\simeq}O(1) \,.
\end{align} 
The above sparseness condition is the analogue of HKS \cite{HKS} sparseness condition, which states that
\begin{align}\label{HKS-sparse}
\log\left(1+\sum_{c/12>\Delta>0}\exp\left(-\beta\Delta\right)\right) \underset{c\to\infty}{\simeq}O(1) \,.
\end{align} 
While the condition, given by the eq.~\eqref{HKS-sparse} is satisfied if and only if the density of states (to be precise, degeneracy $d(\Delta_i)$) for low lying spectra satisfy
\begin{align}
\rho(\Delta) \leq \exp\left(2\pi\Delta\right)\,, \quad \Delta<\frac{c}{12}\,,
\end{align}
the sparseness condition \eqref{sparseness} that we require here is stricter when $\Delta_\chi\neq 0$: 
\begin{align}
\frac{f_{\Delta\mathcal{O}\Delta}}{f_{\chi\mathcal{O}\chi}}\rho(\Delta) \leq \exp\left[2\pi(\Delta-\Delta_\chi)\right]\,, \quad \Delta_\chi<\Delta<\frac{c}{12}\,.
\end{align}
When $\Delta_\chi\neq 0$, we see that our result requires more sparseness in the low lying spectrum at large central charge. The above guarantees that $T(\Delta)$ in large central charge dominated by the lowest nontrivial state with dimension $\Delta_\chi$ only.\\

The rest of the analysis follows in a similar manner for $\delta\sim O(1)$, should we choose 
\begin{align}
\beta=\pi\gamma\sqrt{\frac{1}{3\epsilon}}
\end{align}
and $\Lambda_-$ to satisfy the following condition:
\begin{align}
\Lambda_- \leq 2\pi \sqrt{\left(1-\frac{12\Delta_\chi}{c}\right)\left(1-\frac{1}{12\epsilon}\right)}
\end{align}
This is expected as even in the Cardy formula for $\delta\sim O(1)$, one can naively take the the formula for finite central charge and extend it to infinite central charge \cite{HKS,Baur}. We remark that the reality of $\Lambda_*$ requires $\epsilon>\frac{1}{12}$. Thus $\Delta$ is required to be bigger than $c/6$. This feature is also present in the extended Cardy formula \cite{HKS, Baur}.

\section{Extension and Verification}\label{ex}
The objective of this section is to provide with verification of the results that we derived earlier. Here we work at finite $c$. 
\subsection{Verification I: Identity Module}
In this section we will be considering the operator $\mathcal{Q}\equiv T+\bar{T}$. We remark that $T,\bar{T}$ are $SL(2,R)$ primaries (also known as quasi-primaries); under modular transformation, we have
\begin{align}
\langle\mathcal{Q}\rangle_{\frac{4\pi^2}{\beta}}=-\left(\frac{\beta}{2\pi}\right)^{2}\langle\mathcal{Q}\rangle_{\beta}\,.
\end{align}
Now the lowest state is the identity instead of $\chi$ as we have
\begin{align}
f_{\mathbb{I}\mathcal{Q}\mathbb{I}} = -\frac{c}{12}\,,
\end{align}
and we should set $\Delta_{\chi=\mathbb{I}}=0$. The operator has a spin $2$, thus $i^s f_{\mathbb{I}\mathcal{Q}\mathbb{I}}>0$, hence we can proceed with this operator and conclude that 
\begin{align}\label{exact1}
\mathcal{A} \geq b \left(\Delta-\frac{c}{12}\right)\,.
\end{align}
We emphasize that all the prefactors in the eq.~\eqref{C} involving $c$ cancels out neatly. We know that $\mathcal{Q}$ is an operator corresponding to energy density, thus we have
\begin{align}\label{exact}
f_{\Delta\mathcal{Q}\Delta}=\left(\Delta-\frac{c}{12}\right)
\end{align}
Hence, a direct computation of the quantity $\mathcal{A}$ using Eq.~\eqref{exact} would provide 
\begin{align}\label{exact2}
\mathcal{A} \underset{\Delta\to\infty}{\simeq}  \left(\Delta-\frac{c}{12}\right)
\end{align}
At this point we see that, Eq.~\eqref{exact1} is consistent with Eq.~\eqref{exact2} since $b<1$\footnote{Tighter bounds on $c_-$ ($b$ is related to this $c_-$) in context of Cardy formula is reported in \cite{Ganguly:2019ksp}. Similar functions in principle can be chosen in the present scenario as well with the catch that now the support of the Fourier transform of this functions are different and depends on $\Delta_\chi$.}. This bound on $b$ is guaranteed because of the inequality~\eqref{eq:functionchoice}.\\
%

%

We further point out that one can derive an upper bound using the methodology if  $f_{\Delta\mathcal{O}\Delta}\geq 0$ for all $\Delta>\Delta_*$, where $\Delta_*\sim O(1)$. This happens because then one can write for $\Delta>\Delta_*+\delta$ (this is true eventually as we let $\Delta\to\infty$).
\begin{align}
\nonumber &\int_{\Delta-\delta}^{\Delta+\delta} d\Delta^\prime \sigma(\Delta^\prime)a(\Delta^\prime) e^{-\beta\Delta^\prime}\\
\nonumber &\leq \exp\left[\beta(\Delta+\delta)\right]\int_{\Delta-\delta}^{\Delta+\delta} d\Delta^\prime\ \sigma(\Delta^\prime)a(\Delta^\prime)\ \phi_+(\Delta^\prime)e^{-\beta\Delta^\prime} \\
&\leq  \exp\left[\beta(\Delta+\delta)\right]\int_{\Delta_*}^{\infty} d\Delta^\prime\ \sigma(\Delta^\prime)a(\Delta^\prime)\ \phi_+(\Delta^\prime)e^{-\beta\Delta^\prime} 
\end{align}
where $\phi_+(\Delta^\prime)$ is above the indicator function, as done in \cite{Baur}. 
Now we note that 
\begin{align}
\exp\left[\beta(\Delta+\delta)\right]\int_{0}^{\Delta_*} d\Delta^\prime\ \sigma(\Delta^\prime) a(\Delta^\prime)\ \phi_+(\Delta^\prime) e^{-\beta\Delta^\prime} \underset{\beta\to0}{\simeq} O(1)\times e^{\beta\Delta}=O(1)\times e^{\pi \gamma\sqrt{\frac{c}{3}}} \,,
\end{align}
where we have set $\beta=\pi\gamma\sqrt{\frac{c}{3\Delta}}$. The integral is an order one number, since $\Delta_*$ is an order one number. Thus we find that the term is subleading at large $\Delta$. Henceforth, one can write (we let $\Delta\to\infty$)
\begin{align}
\nonumber &\int_{\Delta-\delta}^{\Delta+\delta} d\Delta^\prime \sigma(\Delta^\prime)a(\Delta^\prime) e^{-\beta\Delta^\prime}\\
&\leq  \exp\left[\beta(\Delta+\delta)\right]\int_{0}^{\infty} d\Delta^\prime\ \sigma(\Delta^\prime) a(\Delta^\prime)\ \phi_+(\Delta^\prime)e^{-\beta\Delta^\prime} 
\end{align}
Now one can proceed in a similar way as done for the lower bound and obtain an upper bound. In fact we know that this is the case for three point coefficients involving quasi-primaries in the identity module, which grows as a polynomial in $\Delta$.\\


\subsection{Verification II: Non Identity module}
\begin{figure}[!ht] 
\centering
\includegraphics[scale=0.5]{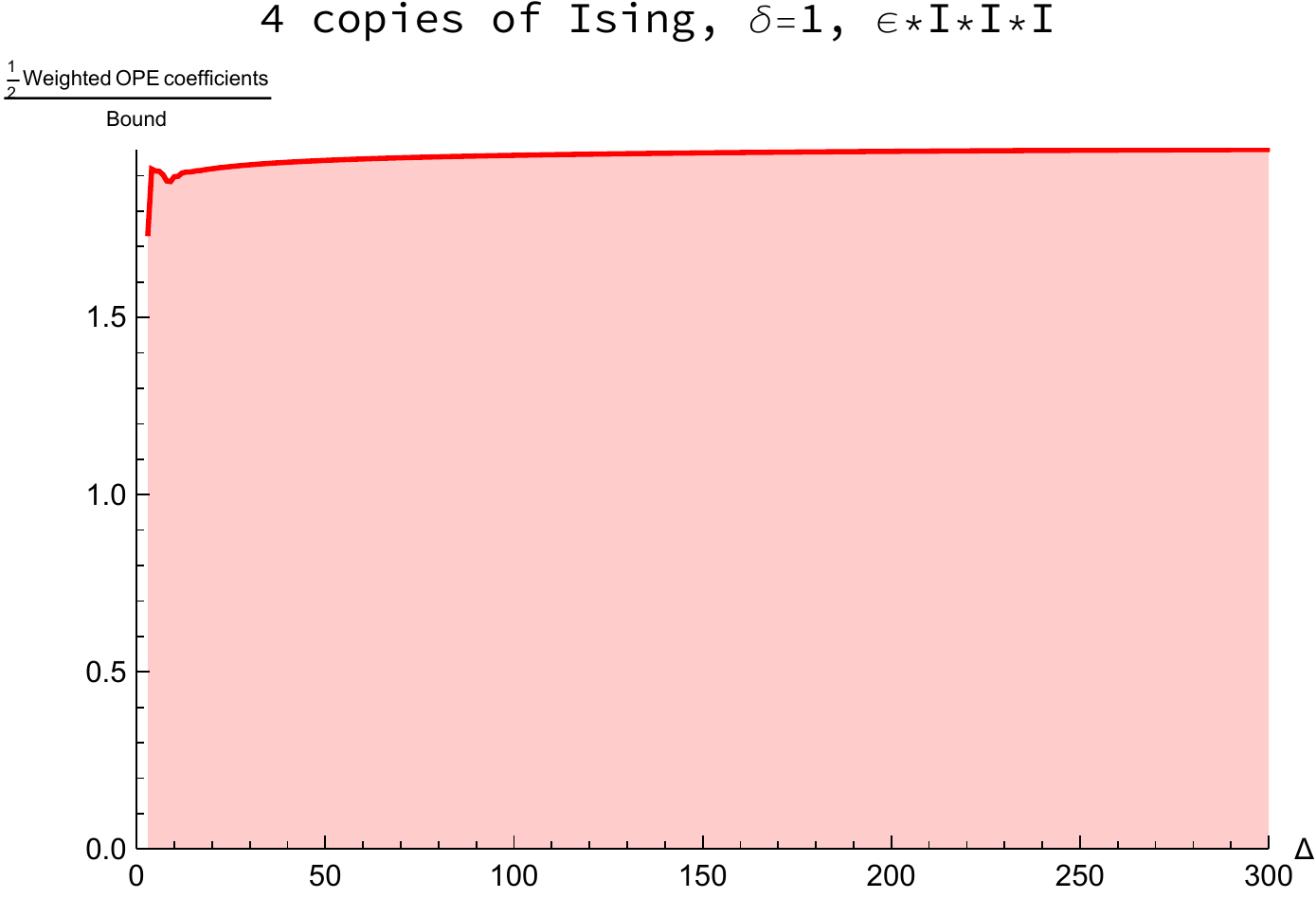}
\includegraphics[scale=0.5]{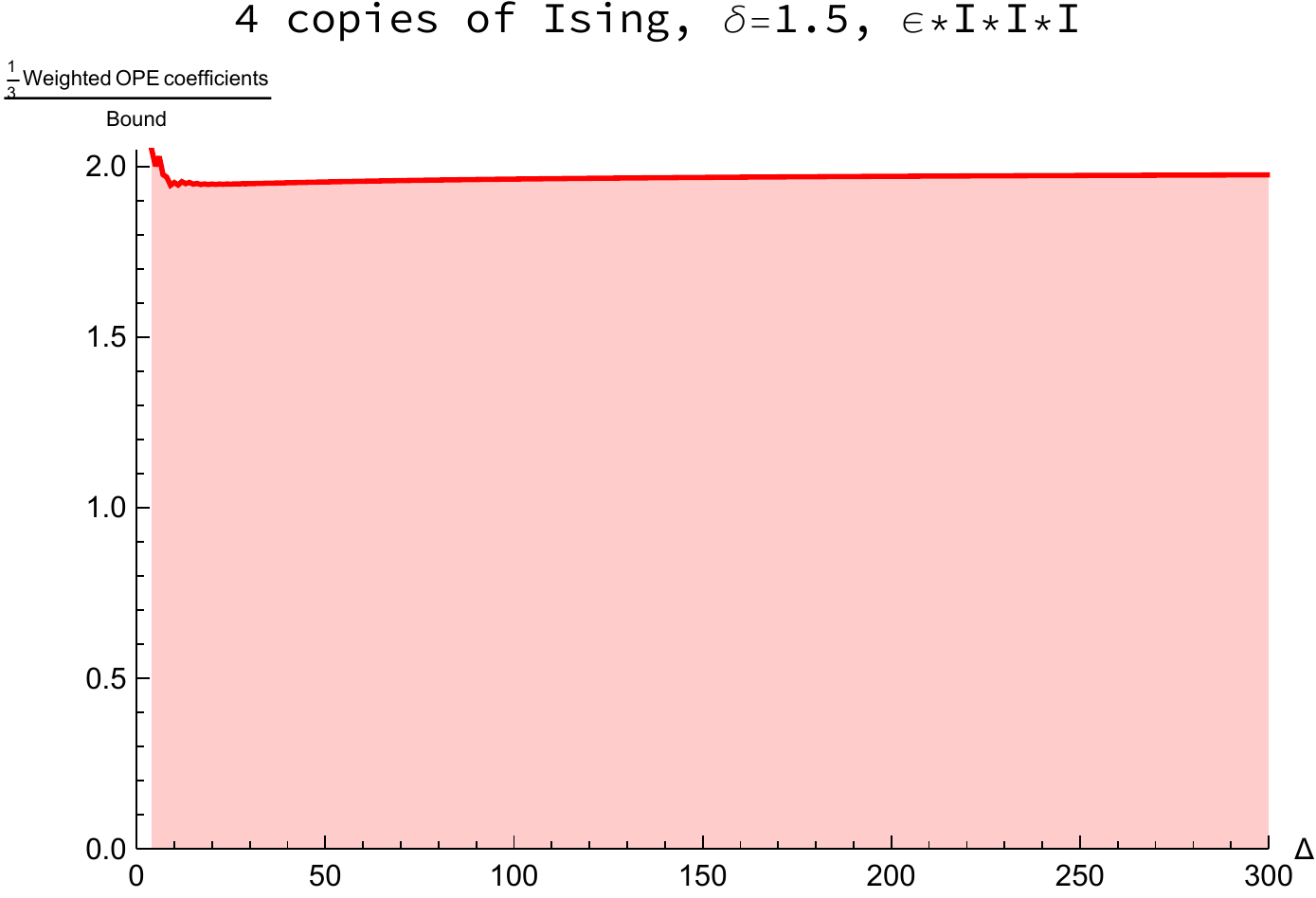}
\caption{Ratio of the weighted three point coefficient and the bound for $4$ copies of Ising model with $c_{eff}=2$, verifying the lower bound. The operator $\mathcal{O}=\epsilon\times\mathbb{I}\times\mathbb{I}\times\mathbb{I}$. We have divided the weighted three point coefficients by the bin width $2\delta$ and we have taken the order one number appearing in the bound to be $0.5$. Technically $0.5$ can only be chosen for $\delta>\frac{1}{\gamma}=2$, for low values of $\delta$, the order one number is less than $0.5$. Nonetheless, if the bound is satisfied with $0.5$, it will be so with any number less than $0.5$.}
\label{check}
\end{figure}

\paragraph{Tensored CFT-I} We consider $4$ copies of $2$D Ising model and tensor them. We take $\mathcal{O}=\epsilon\otimes\mathbb{I}\otimes\mathbb{I}\otimes\mathbb{I}$ with dimension $1$.  The torus one point function of $\mathcal{O}$ is given by
\begin{align}
\langle\mathcal{O}\rangle\propto Z_{\text{Ising}}^{3}\eta^2\,.
\end{align} 
The $q$ expansion of the above can numerically provide us with 
\begin{align}
B=\int_{\Delta-\delta}^{\Delta+\delta}d\Delta^\prime\ \sigma(\Delta^\prime)a(\Delta^\prime)
\end{align}
We have checked for various $\delta$, the above quantity is bounded below by an one number times $T(\Delta)$, as given in the ineq.~\eqref{bound}. The central charge appearing in $T(\Delta)$ should be given by $c_{eff}=2$. In the fig.~\ref{check}, we plot the ratio of $\frac{1}{2\delta}B$ and $0.5\times T(\Delta)$ and verify that it is greater than one number as we vary $\Delta$. Technically $0.5$ can only be chosen for $\delta>\frac{1}{\gamma}=2$, whereas for low values of $\delta$, the order one number is less than $0.5$. Nonetheless, if the bound is satisfied with $0.5$, it will be so with any number less than $0.5$. Since the absolute value is larger, we have $\mathcal{B}\geq B$ and we verify the lower bound as in the ineq.~\eqref{bound} for this model. 

\begin{figure}[!ht]
\centering
\includegraphics[scale=0.5]{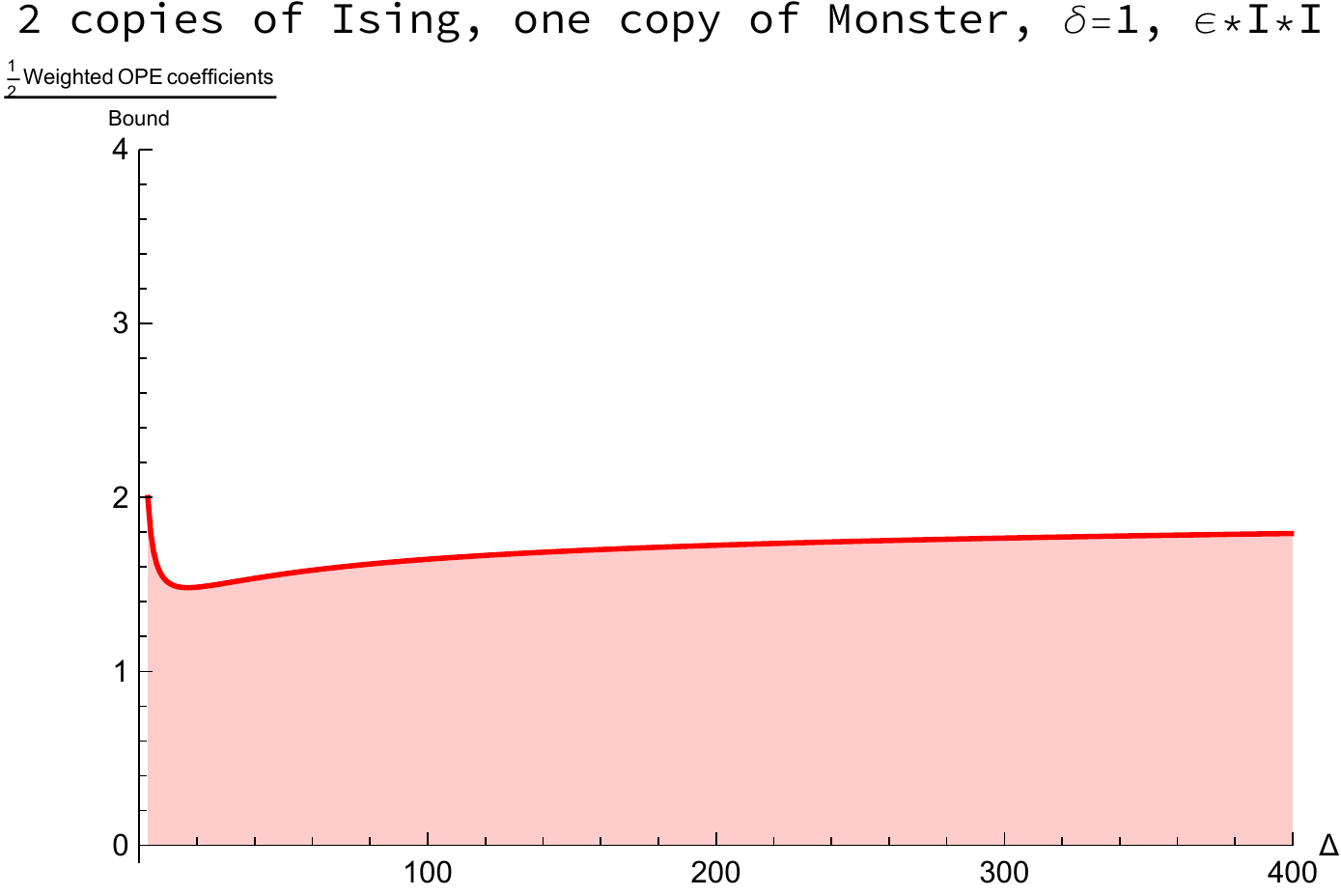}
\includegraphics[scale=0.5]{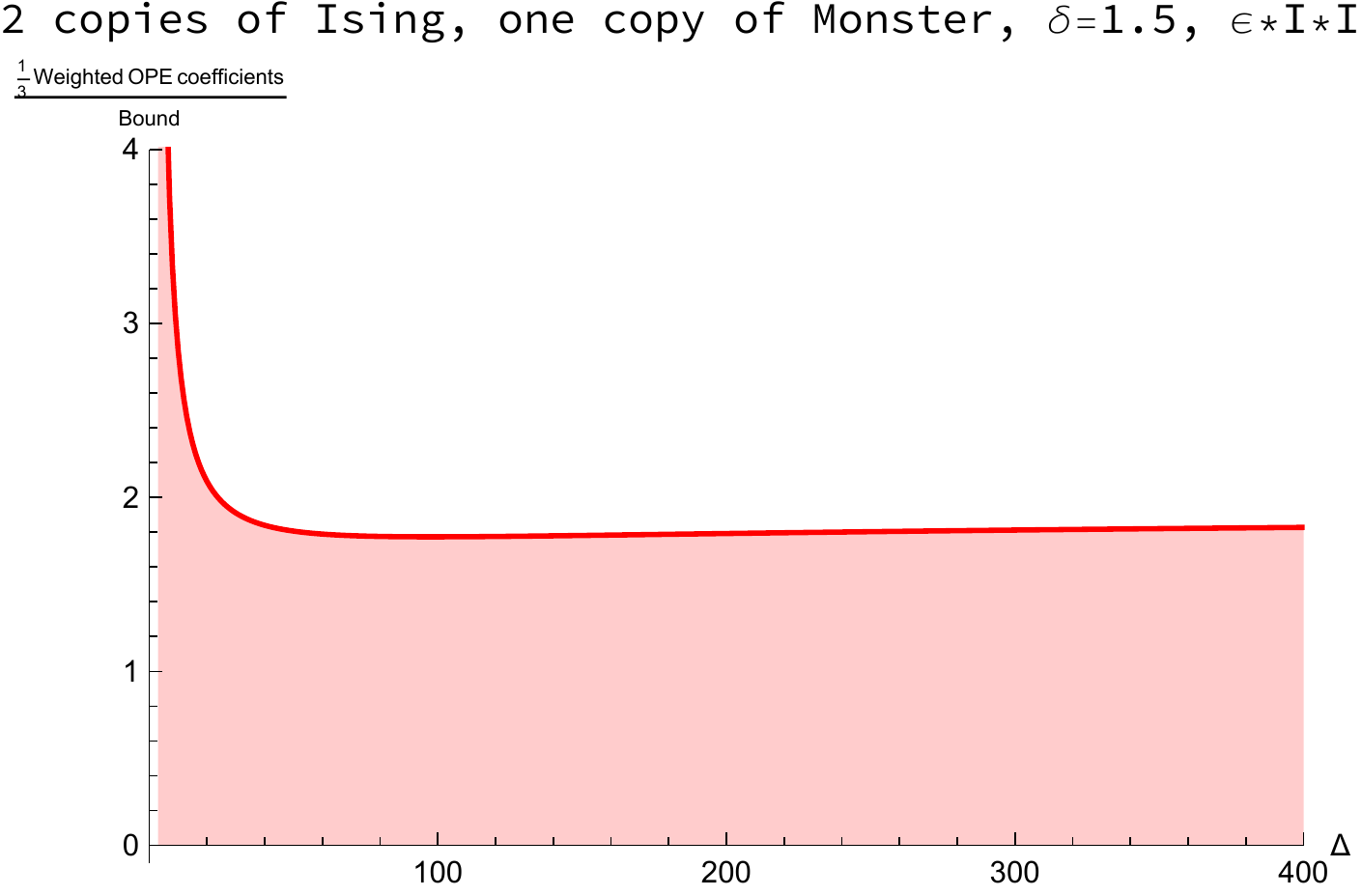}
\caption{Ratio of the weighted three point coefficients and the bound for $2$ copies of Ising model tensored with one copy of Monster such that $c_{eff}=13$. The plot being greater than one  verifyies the lower bound. Here we have divided the weighted three point coefficients by the bin width $2\delta$ and we have taken the order one number appearing in the bound to be $0.5$. Technically $0.5$ can only be chosen for $\delta>\frac{1}{\gamma}\simeq 1.06$, for low values of $\delta$, the order one number is less than $0.5$. Nonetheless, if the bound is satisfied with $0.5$, it will be so with any number less than $0.5$.}
\label{checkmonster}
\end{figure}

\paragraph{Tensored CFT-II}
We consider $2$ copies of $2$D Ising model and one copy of Monster CFT. The tensored CFT has effective central charge of $c_{eff}=13$ (considering the chiral Monster, we have $\frac{c_{eff}}{12}=2\frac{1/2}{12}+\frac{24}{24}$). We look at the operator $\epsilon\otimes\mathbb{I}\otimes\mathbb{I}$. Once 
again we plot the ratio of weighted three point coefficients $\frac{1}{2\delta}B$ and $0.5\times T(\Delta)$ and find that it is bounded below by $1$ for large enough $\Delta$, as can be seen in fig.~\ref{checkmonster}. Here again, technically $0.5$ can only be chosen for $\delta>\frac{1}{\gamma}\simeq 1.06$. For low values of $\delta$, the order one number is less than $0.5$. Nonetheless, if the bound is satisfied with $0.5$, it will be so with any number less than $0.5$.

%
%

\paragraph{Why does it work without the absolute value?} An important remark is in order. In all of the above cases the bound works even without taking the absolute value, this is happening because the $q$ expansion coefficients are what matter for the application of techniques that we have used. Thus even if the individual three point coefficients are negative, it might conspire to sum up to a positive number if there happens to be large number of operators with same dimension $\Delta$. In all of the above case, this is precisely what is happening except possibly for finite number of low lying $\Delta$ as far as we checked numerically. And this could be the potential reason behind why we don't have to take the absolute value for the bound to work. 

\section{The scenario when $\Delta_\chi>\frac{c}{12}$}\label{notkm}
In this section we prove two results pertaining to the case $\Delta_\chi>\frac{c}{12}$. We note that neither the Kraus-Maloney analysis nor the analysis that we did above is applicable when $\Delta_\chi>\frac{c}{12}$. Naively if one tries to extend the result, one would have obtained a power law behavior. Below in the second part, we will see that the qualitative picture of power law behavior is correct.\\

\paragraph{The result I:}The first one states that if $\Delta_\chi>\frac{c}{12}+\frac{\Delta_{\mathcal{O}}}{4\pi}$, then there would always be at least two three point coefficients with opposite sign. This follows from Hellerman \cite{Hellerman:2009bu} like argument applied on the following quantity
\begin{align}
f(\beta)\equiv \left(\frac{\beta}{2\pi}\right)^{\Delta_{\mathcal{O}}/2} \langle \mathcal{O}\rangle_{\beta}
\end{align}

The modular covariance of $\langle \mathcal{O}\rangle_{\beta}$ translates into
\begin{align}
f(\beta)=f\left(\frac{4\pi^2}{\beta}\right)\ \Rightarrow\ \left(\beta\partial_\beta\right)^{N}f(\beta)|_{\beta=2\pi}=0\quad \text{for odd}\ N
\end{align}
The result then follows from considering $N=1$ case\footnote{Another way to state this result would be: if the all three coefficients are positive (or are of same sign), there has to be one operator $\chi$ with dimension below $\frac{c}{12}+\frac{\Delta_{\mathcal{O}}}{4\pi}$ such that $f_{\chi\mathcal{O}\chi}\neq 0$. But we don't know of any example where all the three point coefficients are of same sign.}.\\

To be precise, the above argument implies that there exists $\Delta_1$ and $\Delta_2$ with $\Delta_1\neq \Delta_2$ such that 
\begin{align}
a(\Delta_1)a(\Delta_2)<0
\end{align}

\paragraph{The result II:}The second result states that the $q$ expansion coefficient of $g(s\beta)\equiv \langle\mathcal{O}\rangle_{(s\beta)}$ is of the order of $\Delta^{\Delta_{\mathcal{O}}/2}$ under the following assumptions 
\begin{itemize}
\item $\Delta_\chi>\frac{c}{12}$\\
\item There exists $s\in\mathbb{N}$ such that $\langle\mathcal{O}\rangle_{(s\beta)}$, has an expansion in integer powers of $q$ or it is an expansion of the form $q^{n+\alpha}$, where $n$ is an integer and $\alpha$ is a fixed number. The first one can always be arranged when the set $A$, defined below
\begin{align}
A\equiv \{\Delta: \text{there exists a primary with dimension}\ \Delta\ \&\ \sum_{\Delta_w=\Delta}f_{w\mathcal{O}w}\neq 0\}
\end{align}
is a finite set and $\Delta$ is a rational number.
\end{itemize}
One should contrast this with the estimated $q$ expansion coefficient for the case $\Delta_\chi<c/12$, where we have a exponential growth given by the function in \eqref{defT}.\\

\begin{figure}[!ht]
\centering
\includegraphics[scale=0.7]{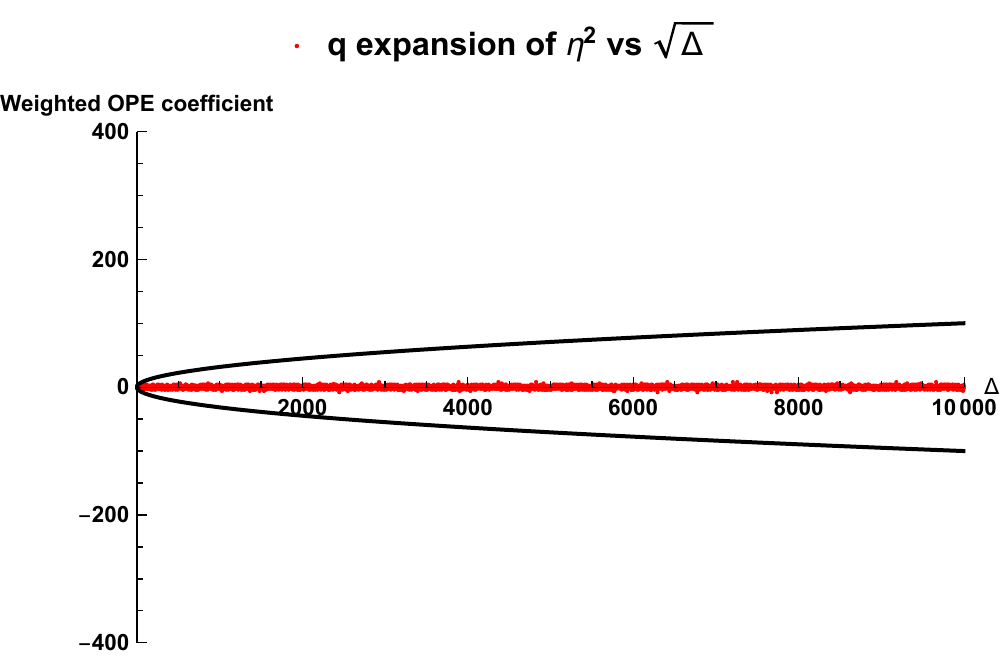}
\includegraphics[scale=0.7]{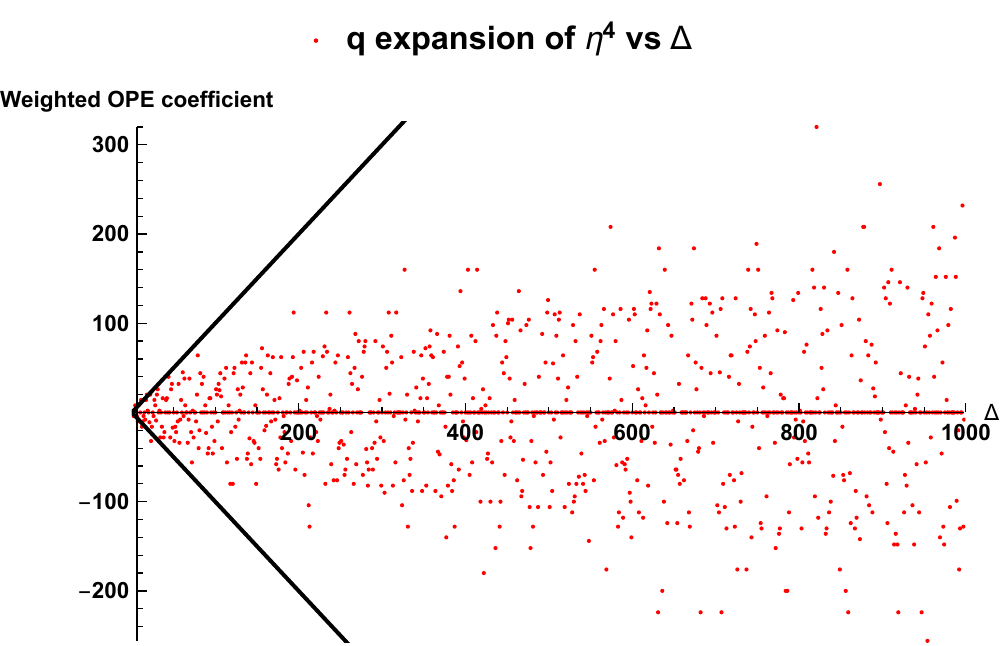}
\caption{The $q$ expansion coefficient as a function of $\Delta$ is bounded by $\Delta^{\Delta_{\mathcal{O}}/2}$, the black lines are $\pm\Delta^{\Delta_{\mathcal{O}}/2}$.}
\label{verifypol1}
\end{figure}

\begin{figure}[!ht]
\centering
\includegraphics[scale=0.7]{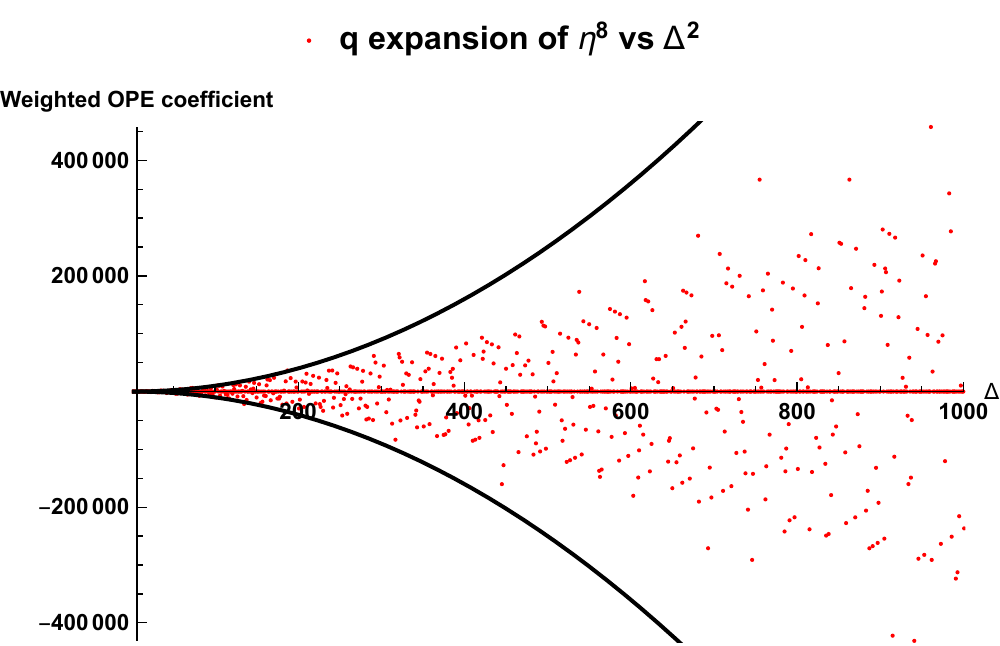}
\includegraphics[scale=0.7]{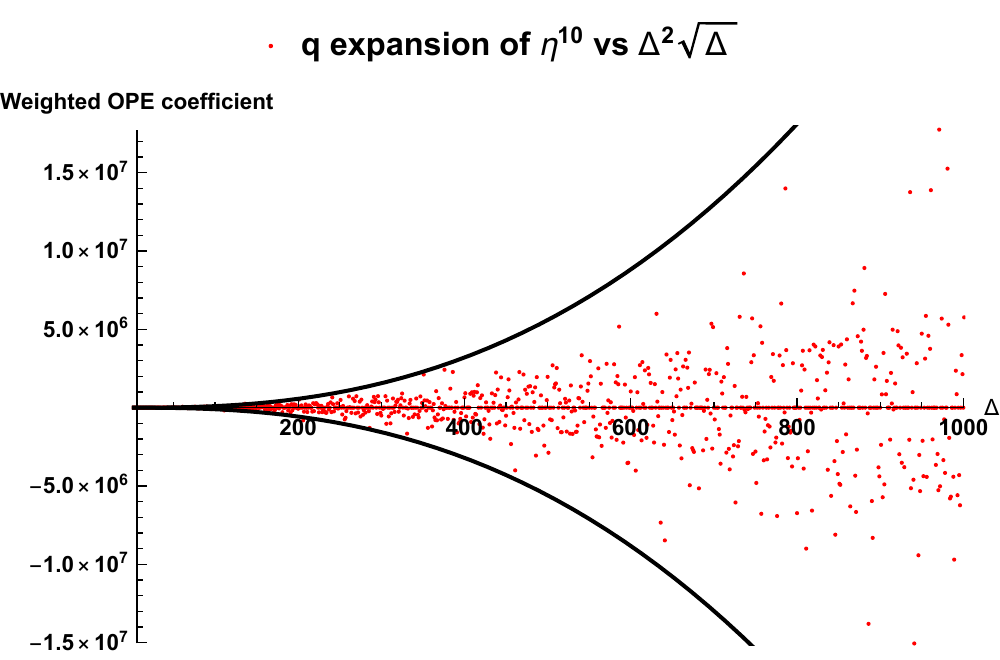}
\caption{The $q$ expansion coefficient as a function of $\Delta$ is bounded by $\Delta^{\Delta_{\mathcal{O}}/2}$, the black lines are $\pm\Delta^{\Delta_{\mathcal{O}}/2}$.}
\label{verifypol2}
\end{figure}

\begin{figure}[!ht]
\centering
\includegraphics[scale=0.7]{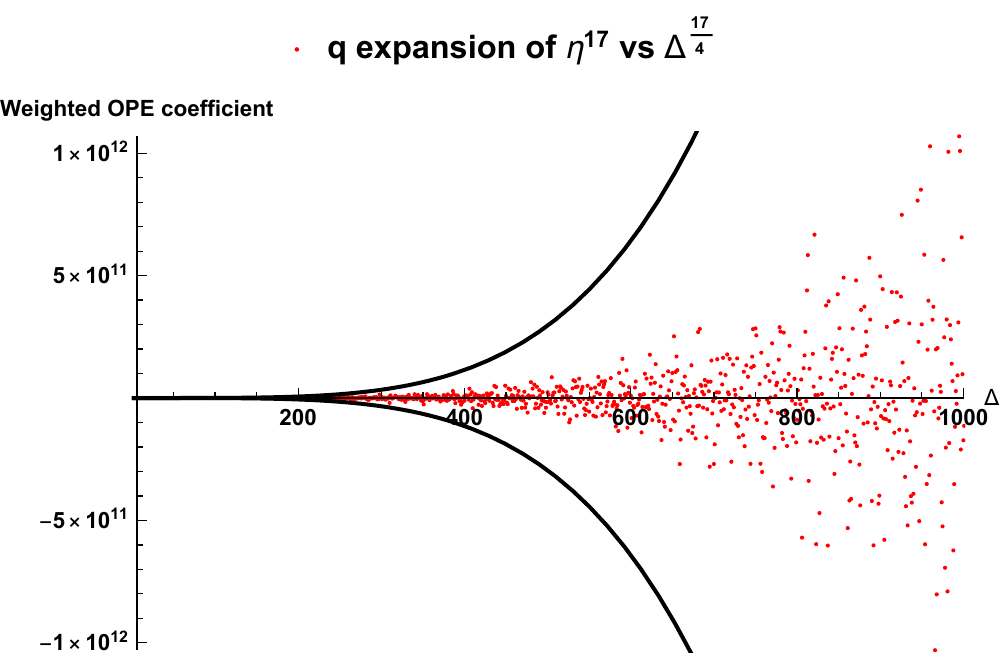}
\includegraphics[scale=0.5]{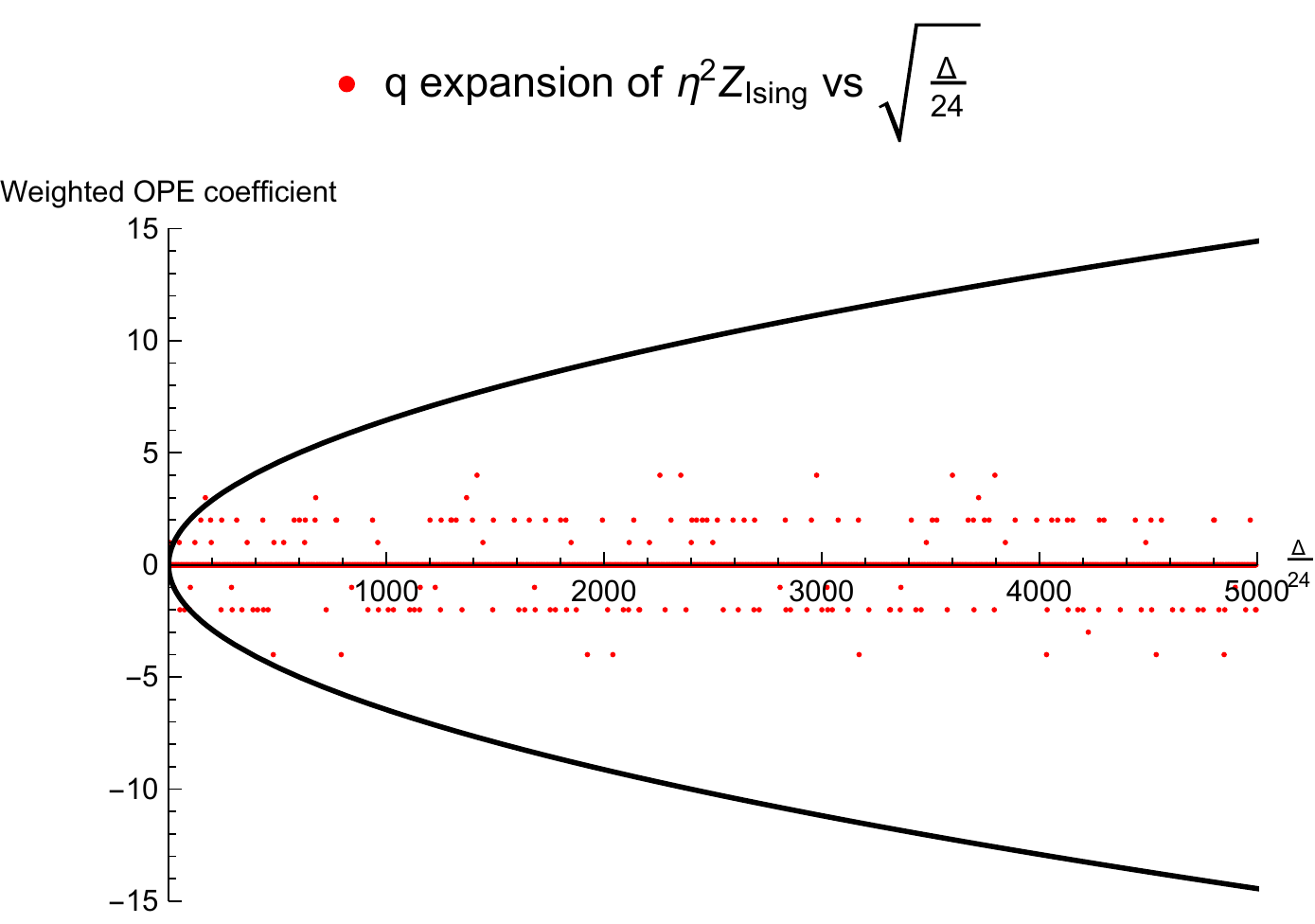}
\caption{The $q$ expansion coefficient as a function of $\Delta$ is bounded by $\Delta^{\Delta_{\mathcal{O}}/2}$, the black lines are $\pm\Delta^{\Delta_{\mathcal{O}}/2}$.}
\label{verifypol}
\end{figure}

In order to prove this let us consider the $q$ expansion of $g(s\beta)\equiv \langle\mathcal{O}\rangle_{(s\beta)}$:
\begin{align}
g(s\beta)\equiv \langle\mathcal{O}\rangle_{(s\beta)}= \sum_n a_{n}q^{n}\,, \quad n\in \mathbb{N}\,
\end{align}
where the fact $n\in \mathbb{N}$ comes from the condition $\Delta_\chi>\frac{c}{12}$. The $q$ expansion coupled with the modular invariance of $f(\beta)$ tells us that 
\begin{align}
|f(\beta+\imath \Omega)|&\leq |f(\beta)| \leq M\\
\Rightarrow |g(\beta+\imath \Omega)| &\leq M \beta^{-\frac{\Delta_{\mathcal{O}}}{2}}
\end{align}
for some order number $M$. Now we use the fact that the powers of $q=\exp\left(-\beta-\imath\Omega\right)$ that appears in the $q$ expansion are intergers, hence we have
\begin{align}
a_{n}=\frac{1}{2\pi\imath}\oint\ \frac{dq}{q^{n+1}}\ g\left(s\beta+\imath s\Omega\right)=\frac{1}{2\pi}\int_{0}^{2\pi}\ d\Omega\ q^{-n} g\left(s\beta+\imath s\Omega\right)\,.
\end{align}
Subsequently one can write
\begin{align}
\nonumber |a_n|&=\frac{1}{2\pi}\bigg|\int_{0}^{2\pi}\ d\Omega\ q^{-n} g\left(s\beta+\imath s\Omega\right)\bigg|\\
&\leq \frac{M}{2\pi} \beta^{-\frac{\Delta_{\mathcal{O}}}{2}}\int_{0}^{2\pi}\ d\Omega\ |q^{-n}| = M \beta^{-\frac{\Delta_{\mathcal{O}}}{2}}\exp\left[-n\beta\right]
\end{align}
Now we choose $\beta=\frac{1}{n}$ to prove that 
\begin{align}\label{exoticbound}
|a_n|\leq M n^{\frac{\Delta_{\mathcal{O}}}{2}} =M s^{\frac{\Delta_{\mathcal{O}}}{2}} \Delta^{\frac{\Delta_{\mathcal{O}}}{2}}
\end{align}
We can verify our results by considering integer powers of $\eta$ function. In fig.~\ref{verifypol1},\ref{verifypol2},\ref{verifypol} we have plotted the $q$ expansion coeffiecient as a function of $\Delta$ to show that they satisfy the bound given in \eqref{exoticbound}.

\begin{figure}[!ht]
\centering
\includegraphics[scale=0.7]{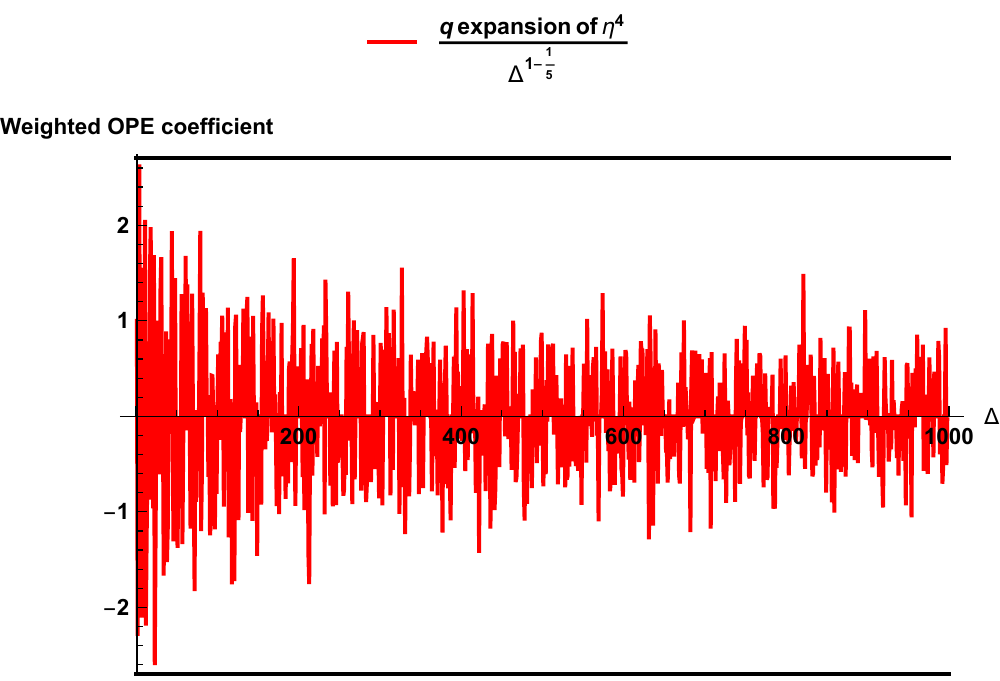}
\includegraphics[scale=0.7]{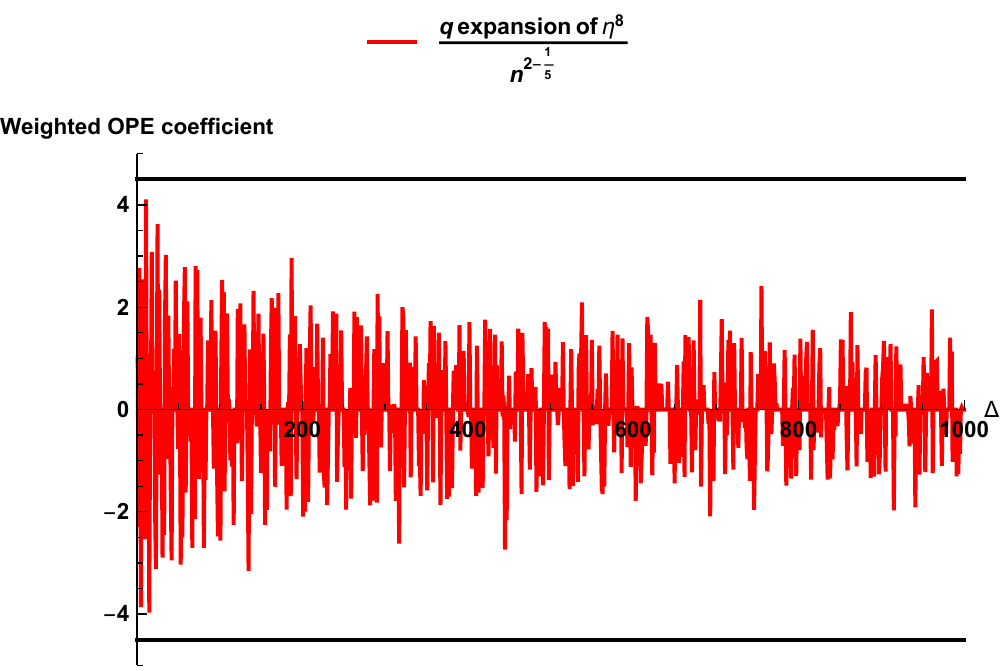}
\caption{The ratio of $q$ expansion coefficient and $\Delta^{\Delta_{\mathcal{O}}/2-1/5}$as a function of $\Delta$ is bounded by an order one number, denoted by the black lines.}
\label{rankinverify}
\end{figure}

\begin{figure}[!ht]
\centering
\includegraphics[scale=0.7]{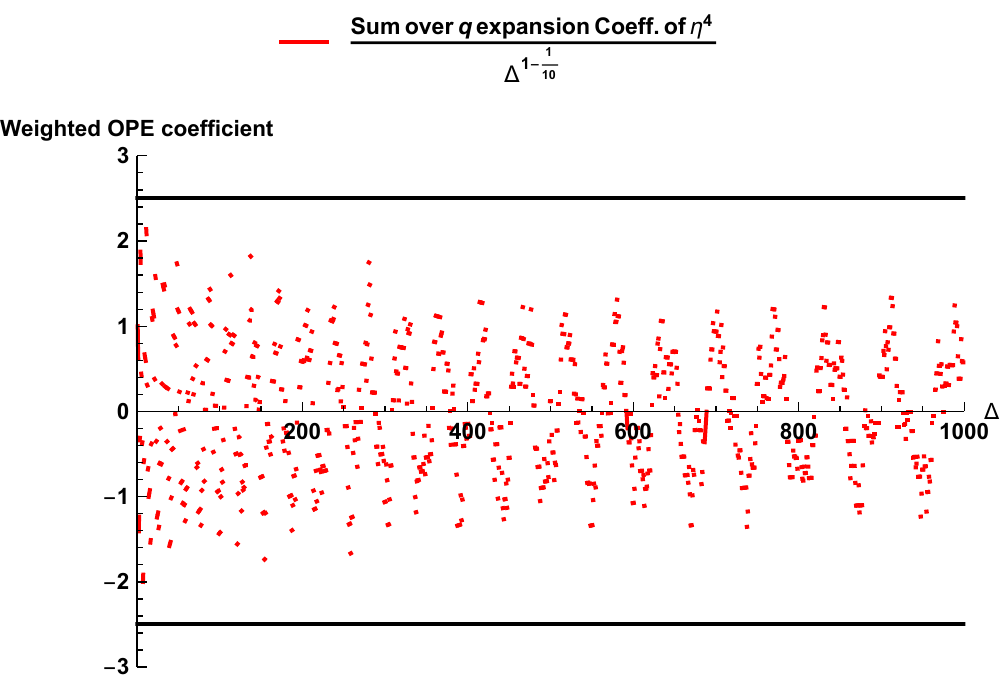}
\includegraphics[scale=0.7]{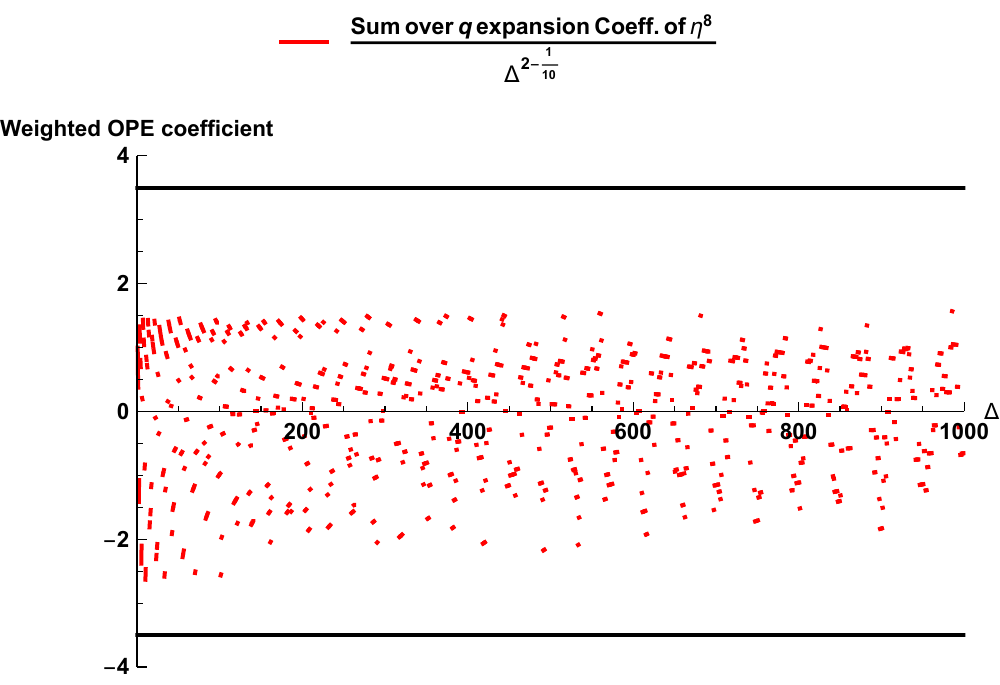}
\caption{The ratio of sum of $q$ expansion coefficient upto $\Delta$ and $\Delta^{\Delta_{\mathcal{O}}/2-1/10}$as a function of $\Delta$ is bounded by an order one number, denoted by the black lines.}
\label{rankinsumverify}
\end{figure}

The analysis above can be made much more precise when $\Delta_{\mathcal{O}}$ is an integer \cite{rankin1939contributions}. Especially theorem 1 and theorem 2 stated there is of direct relevance here. The eq. $1.1$ embodies both the condition mentioned above, which says that 
\begin{align}
H(\tau)=\sum_{n=1}^{\infty}a_n\exp\left[2\pi\imath \tau n/N\right]
\end{align} 
Here $H(\tau)$ is the analogue of one point function. The fact that there is no negative power of $q$ in the expansion indicates $\Delta_\chi>\frac{c}{12}$ while $n/N$ form denotes that in our previous notation, $s=N$. In particular, it is shown that
\begin{align}\label{rankin1}
a_n=O\left(n^{\Delta_{\mathcal{O}}/2-1/5}\right)\,,
\end{align}
which is consistent with what we have shown earlier in this section.  In \cite{rankin1940contributions}, it has further been shown that
\begin{align}\label{rankin2}
\sum_{n\leq X}a_n=O\left(X^{\Delta_{\mathcal{O}}/2-1/10}\right)
\end{align}
Below in the fig.~\ref{rankinverify}, we verify the eq.~\eqref{rankin1} and in the fig.~\ref{rankinsumverify}, we verify the eq.~\eqref{rankin2}.

\paragraph{Tensored CFT and KM result:}Let us reconsider the tensored CFT where we have $4$ copies of $2$D Ising model. If we consider the operator $\mathcal{O}=\epsilon\otimes\mathbb{I}\otimes\mathbb{I}\otimes\mathbb{I}$ and then we have
\begin{align}
f^{\text{tensored CFT}}_{\Delta\mathcal{O}\Delta}=f^{\text{Ising}}_{\Delta_1\epsilon\Delta_1}
\end{align}
Since $\langle\epsilon\rangle=\eta^2$, we know $a_{\Delta_1}=O\left(\Delta_1^{3/10}\right)$, thus we have the typical value of $f^{\text{Ising}}_{\Delta_1\epsilon\Delta_1}$ is suppressed by $P(n)$ , where $n=\Delta_1-\frac{1}{8}$ and $P(n)$ is ``fermionic" partition of integer. Nonetheless, if we want to compute $\mathcal{B}$ we have to integrate over $\left[f^{\text{Ising}}_{\Delta_1\epsilon\Delta_1}\rho(\Delta_1)\prod_{i=2}^{4}\rho(\Delta_i)\right]$ such that $\sum_{i=1}^{4}\Delta_i=\Delta$, which means scanning over a wide range of $f^{\text{Ising}}_{\Delta_1\epsilon\Delta_1}$, only in this integrated form, we can expect the KM result to be valid. This resolves the subtlety associated with the tensored CFT and the KM result \cite{KM}.

\section{Discussion and Outlook} \label{conc}
In this work, we have proved a rigorous lower bound on the asymptotic behavior of the magnitude of the heavy-light-heavy three point coefficients. One might wonder whether considering the negative of the light operator $\mathcal{O}$ would turn the lower bound into an upper bound. First of all, this would not be the case, since we are estimating magnitude of three point coefficients. Second of all, we remark that our prove requires  $f_{\chi\mathcal{O}\chi}(\imath)^{s}>0$ for the lowest dimensional operator $\chi$ contributing to the three point functions. If $f_{\chi\mathcal{O}\chi}(\imath)^{s}<0$, then our results would apply for the operator $-\mathcal{O}$ if the \eqref{modcond} holds true for $-f_{\Delta\mathcal{O}\Delta}$.  We emphasize that among the operator $A_1=\mathcal{O}$ and $A_2=-\mathcal{O}$ we want the condition \eqref{modcond} to be true for the one for which $f_{\chi A_i\chi}(\imath)^{s}>0$. We also proved that if the inequality given in \eqref{modcond} is not satisfied, one can find an increasing subsequence such that the lower bound on the absolute value of $a(\Delta)$ holds, provided we restrict ourselves to the subsequence while taking $\Delta\to\infty$ limit. We have also shown that an upper bound can be obtained if the three point coefficients are positive except for a finite number of $\Delta$. Some thoughts over proving the upper bound in a more generic set up is presented in appendix~\S\ref{app:A}. We hope to fill in the details in future.\\

The result continues to hold at a large central charge under a certain sparseness condition on the spectra. The sparseness condition, as it turns out, is stronger than the sparseness condition \cite{HKS} required for Cardy like behavior of the density of states at large $c$ with $\Delta/c$ being finite. Furthermore, if one does not assume that $\Delta_\chi\sim O(c)$, the three point coefficients do not fall off exponentially in central charge.\\ 


In this paper we have considered operators $\mathcal{O}$ with even spin. If we only have integer spin operators with dimension $\Delta$ in the spectrum, then $f_{\Delta\mathcal{O}\Delta}=0$ if $\mathcal{O}$ has odd spin, this motivates our choice of assuming that $\mathcal{O}$ has even spin. We remark that the partition function is not invariant under $T$ modular transformation if we have operators with half-integer spin. Nonetheless, if one considers $S$ modular transformation only, it is possible to have operators with half integer spin, in that scenario, $\mathcal{O}$ can possibly carry odd spin with $f_{\Delta\mathcal{O}\Delta}$ being non-zero, in which case, the operator with dimension $\Delta$ must carry half integer spin. The odd spin introduces a factor of $i$ in the analysis, but this gets compensated precisely by the same factor appearing in $f_{\chi\mathcal{O}\chi}$\footnote{The author thanks Ken Intriligator for discussion on this point.} \cite{Albayrak:2019gnz}.\\

We have verified our result for the quasi-primaries belonging to the Identity module and for nontrivial primary involving $\epsilon$ operator of $2$D Ising model. It is not clear whether similar bound can rigorously be obtained specific for heavy primaries. This would require knowledge about the torus conformal block. This might be tractable at large central charge. Another obvious extension of this work is to consider extended Virasoro algebra and putting the results of \cite{dattadaspal} on rigorous footing. It would also be interesting to investigate the torus two point function \cite{Brehm:2018ipf,Romero-Bermudez:2018dim,Hikida:2018khg} using similar methods. \\

We have shown that in the large central charge limit, the average value  $\mathcal{A}\geq \langle\mathcal{O}\rangle_{\beta_{ETH}}$, this raises the question of whether the relative sign of three point coefficients is important to match up with the ETH prediction, since ETH predicts the behavior of three point coefficients whereas we are probing the absolute value of the three point coefficients for heavy states.\\

\begin{figure}[!ht]
\centering
\includegraphics[scale=0.7]{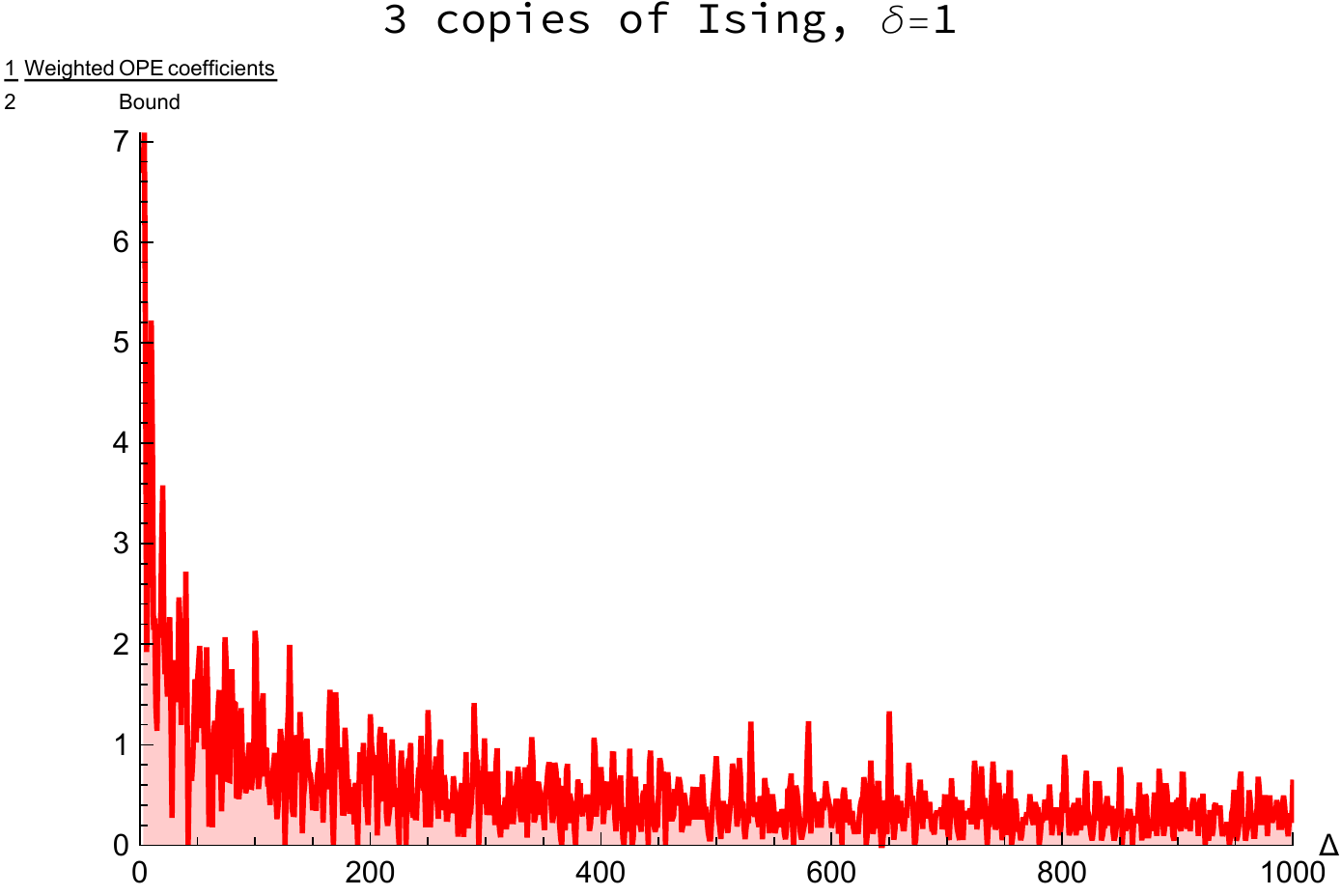}
\caption{The ratio of $\frac{1}{2\delta}B$ and $\frac{1}{\sqrt{2}}(\Delta-\frac{1}{8})^{1/2}$ for 3 copies of Ising model with $\delta=1$. The operator $\mathcal{O}=\epsilon\otimes\mathbb{I}\otimes\mathbb{I}$.}
\label{boundarycase}
\end{figure}

We have derived the order of magnitude of $q$ expansion coefficients of one point function when $\Delta_\chi>\frac{c}{12}$. This behavior is sharply different from the case where $\Delta_\chi<\frac{c}{12}$. In the former, we have polynomial growth while in the later we have exponential growth of $q$ expansion coefficient. We remark that the case when $\Delta_\chi=\frac{c}{12}$ is still open, since neither method is applicable in this scenario. On the other hand, we do have examples where this happens, for example consider $3$ copies of $2$D Ising model with $c_{eff}=\frac{3}{2}$ and $\mathcal{O}=\epsilon\otimes\mathbb{I}\otimes\mathbb{I}$ with $\Delta_\chi=\Delta_{\sigma\times\mathbb{I}\times\mathbb{I}}=\frac{1}{8}$. It would be interesting to fill in this gap. Naively one would expect power law behavior for $a(\Delta)$ in such scenario. In fact a plot of the ratio of $\frac{1}{2\delta}B$ and $\frac{1}{\sqrt{2}}(\Delta-\frac{1}{8})^{1/2}$ looks as in the fig.~\ref{boundarycase}.
\\

We end with a paragraph speculative in nature. If the inequality~\eqref{modcond} is satisfied, then it is tempting to conjecture an existential result. In particular, one can hope to show that there exists a subsequence of $\Delta_{k}$s such that $\mathcal{B}$ for a small interval centered around $\Delta_k$ asymptotes to the function $T(\Delta)$ (which earlier appeared as a lower bound, so we need to prove an upper bound to show the $T(\Delta)$ is the asymptotic bheavior) with an error that goes like polynomial or one over a polynomial. This can possibly be done in following way. If there is only finite number of negative $a(\Delta)$, we can prove an upper bound as outlined earlier in section where we discuss $T$. If there is an infinite number of negative $a(\Delta)$, then we form a sequence $\Delta_k$ such that magnitude of each of the term in the sequence $a(\Delta_k)$ is upper bounded. One can hope to extend this to $\mathcal{B}$ for small energy window centered around $\Delta_k$ (this is the missing link right now). This can for example be done by making the energy window $\delta$ small enough via suitable choice of a function $\phi_-$. We refer to \cite{Ganguly:2019ksp} for the optimal $\phi$, but it would cost us a relative error suppressed by $\Delta^{-y}$ for some positive number $y$. \\

The second speculation is regarding a curious observation that we make: in all of the examples involving tensored CFTs, where $\Delta_\chi<c/12$, the $q$ expansion coefficient turns out to be positive except possibly for a finite number of $\Delta$. If this is really a generic feature and can be proven, then one can simply get rid of  the eq.~\eqref{modcond} and find asymptotics of three point coefficients without considering the absolute value.

\section*{Acknowledgements}
The author thanks Baur Mukhametzhanov and Alexander Zhiboedov for helpful suggestions and comments and asking whether minimal models satisfy the bound. The author acknowledges Ken Intriligator, John McGreevy, Luis Fernando Alday for fruitful discussions and encouragement. The author  thanks Diptarka C Das and Eric Perlmutter for going through the initial draft and providing helpful comments. The author also thanks Shouvik Datta for comments. This work was in part supported by the US Department of Energy (DOE) under cooperative research agreement DE-SC0009919 and Simons Foundation award  \#568420. The author also acknowledges the support from Inamori Fellowship, Ambrose Monell Foundation and DOE grant DE-SC0009988. 

\section*{An anecdote}
I have been curious about the applicability of Kraus-Maloney result to the tensored CFT since 2017 when I started working on extending the result for extended Virasoro algebra with Diptarka Das and Shouvik Datta. The issue was brought up then by John McGreevy in a different context. The initial guess was that the tensored CFT has more symmetry than Virasoro, hence KM would not be applicable. Nonetheless, nowhere in KM's analysis had it ever mentioned that the only symmetry algebra is Virasoro. Hence the issue was not really resolved and I visited the problem on and off. This work happens to resolve this paradox by carefully looking at the window over which the averaging is being done. 

\appendix
\section{More on Tauberian theorems \& upper bound}\label{app:A}
In general, Tauberian theorems require positivity, for example the famous Hardy-Littlewood theorem (it was first proven by Hardy and Littlewood, later a simplified proof was given by Karamata.). It states:\\

If $a_n\geq 0$ and
\begin{align}
f(x)&=\sum_n a_nx^n \sim \frac{C}{1-x}\,, \quad x\to 1
\end{align}
the following holds true:
\begin{align}
\sum_{k=0}^{N} a_k \sim CN\,,\quad N\to\infty
\end{align}

Now the positivity condition can be relaxed to the following:
\begin{align}\label{tcond}
a_n\geq -H\,,\quad H\sim O(1)\,.
\end{align}
One can easily see why this is possible: we consider an auxiliary sequence defined by 
\begin{align}
b_n=a_n+H\geq 0
\end{align}
such that we have
\begin{align}
\sum_n b_nx^n \sim \frac{C+H}{1-x}\,,\quad x\to 1
\end{align}
Now applying the original theorem with the requirement of positivity, we obtain 
\begin{align}
\sum_{k=0}^{N}b_k& \sim (C+H)N\,,\quad N\to\infty\,, \\
\Rightarrow \sum_{k=0}^{N}a_k &\sim CN\,,\quad N\to\infty\,.
\end{align}

A more detailed exposition of the above can be found in the chapter 7 of the book \cite{hardy2000divergent} by G.H.Hardy.\\ 

Below we provide the readers with a sketch of how a proof of the above kind would look like. We hope to fill in the details in future. In stead of individual heavy light heavy three point coefficients, we consider $a(\Delta)$, which is in fact the sum of all the heavy light heavy three point coefficients for a particular $\Delta$. In other words, we have
\begin{align}
\langle\mathcal{O}\rangle_{\beta}=\sum_{\Delta^\prime}a(\Delta^\prime)e^{-\beta\left(\Delta^\prime-\frac{c}{12}\right)}
\end{align} 
In order to sketch the method of our proof, we consider the following condition:
\begin{align}\label{eq:tauberianconditon}
a(\Delta^\prime)>-\exp\left(-\alpha\Delta^\prime\right)\,,\quad \alpha\sim O(1)\,.
\end{align}
We should be thinking of the above as analogue of the eq.~\eqref{tcond}. Now one can define 
\begin{align}
b(\Delta^\prime)=a(\Delta^\prime)+\exp\left(-\alpha\Delta^\prime\right)
\end{align}
In $\Delta\to\infty$ limit, we indeed have 
\begin{align}
\int_{\Delta-\delta}^{\Delta+\delta} b(\Delta^\prime)\sigma(\Delta^\prime)\sim \int_{\Delta-\delta}^{\Delta+\delta} a(\Delta^\prime)\sigma(\Delta^\prime)
\end{align}

We further note that the extra term $\exp\left(-\alpha\Delta^\prime\right)
$ does not alter the behavior of the integrated three point coefficient by that much, since we have
\begin{align}
\left|\Sigma_{\pm}\right| \leq Ke^{\pi\gamma\sqrt{\frac{c\Delta}{3}}}=\left[\rho_0(\Delta)\right]^{\gamma/2}\,,\quad K\sim O(1)\,,
\end{align}
which is subleading to $T(\Delta)$, as defined in the eq.~\eqref{defT}. Here we have
\begin{align}
\Sigma_{\pm}(\beta)\equiv e^{\beta\left(\Delta\pm\delta\right)} \int_{0}^{\infty} d\Delta^\prime\ \exp\left[-(\alpha+\beta)\Delta^\prime\right]\sigma(\Delta^\prime)\phi_{\pm}(\Delta^\prime) 
\end{align}
and we recall that $\phi_\pm$ approximates the indicator function from the above and the below and $|\phi_\pm|$ is bounded by an order one number.\\

On the other hand, since $b(\Delta)$ is positive definite, the analysis of asymptotics of $b(\Delta)$ falls naturally under the umbrella of the Tauberian theorems. In particular, one can set up an inequality of the following form 
\begin{align}
\nonumber &e^{\beta\left(\Delta-\delta\right)}\int_{0}^{\infty}d\Delta^\prime\ b(\Delta^\prime)\sigma(\Delta^\prime)\phi_-(\Delta^\prime)e^{-\beta\Delta^\prime} \\ \nonumber&\leq\int_{\Delta-\delta}^{\Delta+\delta}d\Delta^\prime\ b(\Delta^\prime)\sigma(\Delta^\prime) \\ & \leq e^{\beta\left(\Delta+\delta\right)}\int_{0}^{\infty}d\Delta^\prime\ b(\Delta^\prime)\sigma(\Delta^\prime)\phi_+(\Delta^\prime)e^{-\beta\Delta^\prime}
\end{align}
and proceed using the method described in the main text. Note we have already argued that the extra bit in $b(\Delta)$ contributes in a subleading manner. Thus one can arrive at a lower bound as well as an upper bound for $\mathcal{B}$. The examples provided in the main text do satisfy the constraints of theorem and one can see that both the lower and upper bound is satisfied in fig.~\ref{check} and \ref{checkmonster} since the curve asymptotes/saturates to a constant value. For the fig.~\ref{check40}, one might need to probe high enough value of $\Delta$ to see the saturation. As mentioned briefly in the main text at the end of \S\ref{M}, the above kind of argument can indeed be applied coupled with Ingham's theorem and thus would lead to a true asymptotic estimate of $a(\Delta)$.

\section{One more example!}\label{app:B}
\begin{figure}[!ht]
\centering
\includegraphics[scale=0.7]{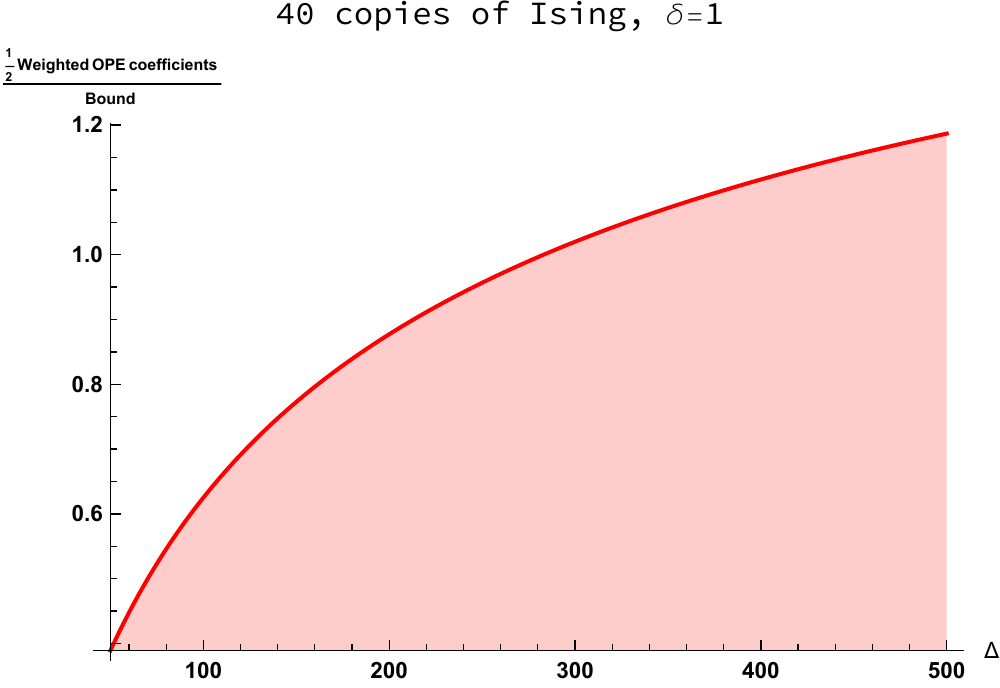}
\caption{Ratio of the weighted three point coefficients and the bound for $40$ copies of Ising model with $c_{eff}=20$, verifying the lower bound since the curve eventually goes above $1$. Here we have $\mathcal{O}\equiv \left(\otimes^{12}\epsilon\right)\left(\otimes^{28}\mathbb{I}\right)$}
\label{check40}
\end{figure}

We reconsider the example of tensoring $40$ copies of $2$D Ising model and consider the operator $\mathcal{O}\equiv \left(\otimes^{12}\epsilon\right)\left(\otimes^{28}\mathbb{I}\right)$. Here also, we check that the ratio of weighted three point coefficients integrated over an integral and the bound is an increasing function of $\Delta$, as can be seen in the fig.~\ref{check40}.

{\bibliographystyle{bibstyle2017}
\bibliography{refs}
\hypersetup{urlcolor=RoyalBlue!60!black}
}

\end{document}